\documentclass[conference]{IEEEtran}
\usepackage{cite}
\usepackage{amsmath,amssymb,amsfonts}
\usepackage{algorithmic}
\usepackage{graphicx}
\usepackage{comment}
\usepackage{textcomp}
\usepackage[hyphens]{url}

\usepackage[dvipsnames]{xcolor}
\usepackage{soul}
\usepackage{hyperref}
\hypersetup{
  breaklinks=true,
  colorlinks=true,
  allcolors=BlueViolet,
}

\def\BibTeX{{\rm B\kern-.05em{\sc i\kern-.025em b}\kern-.08em
    T\kern-.1667em\lower.7ex\hbox{E}\kern-.125emX}}

\usepackage{braket}

\title{Clifford-based Circuit Cutting for \\ Quantum Simulation}

\author{\IEEEauthorblockN{Kaitlin N. Smith\textsuperscript{\textsection}, Michael A. Perlin\textsuperscript{\textsection}, Pranav Gokhale, Paige Frederick, \\David Owusu-Antwi, Richard Rines, Victory Omole, and Frederic T. Chong}
\IEEEauthorblockA{\textit{Super.tech, a division of Infleqtion}\\ 
}
}

\newcommand{\old}[1]{}

\begin{document}

\maketitle
\begingroup\renewcommand\thefootnote{\textsection}
\footnotetext{These authors contributed equally.}
\endgroup

\maketitle
\thispagestyle{plain}
\pagestyle{plain}


\begin{abstract}

Quantum computing has potential to provide exponential speedups over classical computing for many important applications. However, today's quantum computers are in their early stages, and hardware quality issues hinder the scale of program execution.
Benchmarking and simulation of quantum circuits on classical computers is therefore essential to advance the understanding of how quantum computers and programs operate, enabling both algorithm discovery that leads to high-impact quantum computation and engineering improvements that deliver to more powerful quantum systems. Unfortunately, the nature of quantum information causes simulation complexity to scale exponentially with problem size. 

In this paper, we debut Super.tech's SuperSim framework, a new approach for high fidelity and scalable quantum circuit simulation. SuperSim employs two key techniques for accelerated quantum circuit simulation: Clifford-based simulation and circuit cutting. Through the isolation of Clifford subcircuit fragments within a larger non-Clifford circuit, resource-efficient Clifford simulation can be invoked, leading to significant reductions in runtime. After fragments are independently executed, circuit cutting and recombination procedures allow the final output of the original circuit to be reconstructed from fragment execution results. Through the combination of these two state-of-art techniques, SuperSim is a product for quantum practitioners that
allows quantum circuit evaluation to scale beyond the frontiers of current 
simulators. Our results show that Clifford-based circuit cutting accelerates the simulation of near-Clifford circuits, allowing 100s of qubits to be evaluated with modest runtimes.

\end{abstract}

\section{Introduction}
Quantum computing is an emerging information processing paradigm that shows great theoretical promise for applications such as chemistry~\cite{kandala2017hardware}, optimization~\cite{moll2018quantum}, cryptography~\cite{shor1999polynomial}, and machine learning~\cite{biamonte2017quantum}, among others. Because quantum bits, or qubits, have the ability to demonstrate quantum superposition, interference, and entanglement, quantum algorithms enable significant speedups when applied toward certain classes of problems, usually those characterized by a large search space in which the ``optimal'' solution lives. 

Quantum circuit simulation with classical hardware offers an effective means to quantify performance and troubleshoot potential issues within quantum programs without needing direct access to quantum hardware. Through quantum circuit simulation, the computational power of quantum and classical computers can be distinguished, enabling the discovery of applications that have non-trivial quantum advantage. Additionally, quantum circuit simulators that faithfully reflect real machine noise are a viable pathway to identify quantum computer (QC) hardware features critical to the success of quantum algorithms, leading to engineering improvements that deliver to more powerful quantum systems. Unfortunately, state-of-art classical methods for quantum circuit simulation suffer from scaling challenges: resource requirements grow exponentially with the size of the quantum problem.
While this inability to easily translate many quantum computations into a classical equivalent enables QCs to demonstrate advantage when applied toward certain domains, it also hinders the understanding of large-scale quantum processing.
Fortunately, not all classical simulation of quantum circuits is characterized  by intractable scaling. If a circuit is expressed with a special family of operations, known as Clifford operations, quantum circuit simulation becomes efficient on classical hardware, with a classical simulation complexity that grows quadratically with quantum system size.

In this industry-track paper, we describe Super.tech's development of SuperSim, the first simulator to unite two techniques: Clifford circuit simulation and circuit cutting. \emph{Clifford circuit simulation} is the efficient simulation of an important class of quantum circuits found in many critical quantum applications. \emph{Circuit cutting} is a divide-and-conquer framework for simulating large quantum circuits by cutting them into smaller subcircuits that can be independently executed on smaller QCs. By uniting these techniques, SuperSim expands the set of quantum applications for which classical simulation is tractable and helpful.

The class of circuits that SuperSim addresses is referred to as \textit{near-Clifford} circuits. Prior work has also addressed near-Clifford circuits. However, our approach is the first to leverage the recent development of quantum circuit cutting. We also benefit from the recent emergence of new applications of near-Clifford circuits to solving chemistry~\cite{ravi2022cafqa} and optimization~\cite{medvidovic2021classical} problems. 

SuperSim is not a panacea for simulating near-Clifford circuits. Although our specific setting has some computationally advantageous factors over generic circuit cutting (see Section~\ref{sec:clifford_specific_cutting_optimizations}), SuperSim still shares the limitations of generic circuit cutting. Specifically, if a circuit requires many ($k$) cuts to partition into Clifford and non-Clifford subcircuits, the reconstruction time (scaling as $4^k$) can be intractable. Despite this asymptotic limitation in the limit of a large number of cuts, we show in Section~\ref{sec:results} that in two example applications, SuperSim offers orders-of-magnitude faster simulation over other state-of-the-art approaches. We hypothesize that additional near-Clifford quantum circuits with potential for improved classical simulation via SuperSim have yet to be discovered.
These advantages are apparent even in a straightforward, naive implementation of SuperSim; we discuss how the performance of SuperSim can be further improved by leveraging parallelization and GPU acceleration in Section~\ref{sec:ongoing-performance-improvements}.

The launch of SuperSim will encompass the release of the open-source codebase, as well as accompanying documentation, tutorials, and product framework that are currently under development. Our hope is that SuperSim will be an invaluable tool for architects and quantum practitioners, who will apply it to study near-Clifford circuits pertinent to error correction design and to end-user applications. Moreover, we will release SuperSim under an open-source license to encourage developers to continuously improve the codebase and expand SuperSim’s applications. From an industry standpoint, this open-source model aligns with our strategic goals -- we anticipate commercial benefit if and when users go beyond local simulators to running on classical accelerators (i.e. GPUs, TPUs) and real quantum hardware.

This paper provides an evaluation of the SuperSim framework: a new software solution that accelerates the classical simulation of near-Clifford quantum circuits. As SuperSim is in its alpha version, we plan to release upcoming improvements in terms of infrastructure, verification, and benchmarking that are part of Super.tech's product road map. The remainder of this paper is structured as follows. Section~\ref{QC-fundamentals} provides background information on quantum computation that aids paper comprehension. Section~\ref{sec:prior-work-mot} motivates the development of the SuperSim framework by describing the limitations involved with the classical simulation of quantum circuits, the benefits of Clifford simulation, and the basics of quantum circuit cutting. Section~\ref{sec:applications-of-near-clifford-simulation} explains the importance of near-Clifford circuits and provides example applications for their simulation. Section~\ref{sec:SuperSim-framework-overview} introduces SuperSim itself\old{: the Super.tech framework that invokes Clifford-based circuit cutting for quantum circuit simulation}. Section~\ref{sec:methods} describes our methodology to evaluate SuperSim when targeted toward select applications described in Section~\ref{sec:applications-of-near-clifford-simulation}. Section~\ref{sec:discussion} includes a discussion of the implications of this paper's preliminary SuperSim evaluations as well as related work. Section~\ref{sec:clifford_specific_cutting_optimizations} discusses in-progress framework optimizations that are specific to SuperSim's Clifford-based cutting techniques. Section~\ref{sec:ongoing-performance-improvements} presents other general SuperSim performance improvements outlined in our development roadmap. Section~\ref{sec:ongoing-roadmap-for-future-eval} describes lessons learned and how we plan to improve the evaluation of SuperSim. Finally, Section~\ref{sec:conclusion} summarizes this work's findings and provides conclusions.

\section{Quantum Computing Fundamentals}
\label{QC-fundamentals}

This section contains an overview of quantum computing fundamentals that are essential to our study. The subsections here are not all-inclusive. The authors encourage the curious reader to find a more detailed discussion of the following topics in Ref.~\cite{mike_ike_2020}.

\subsection{Quantum Information}
\label{information-qc}

An isolated qubit can hold a complex-valued linear combination (\emph{superposition}) of the basis states $\ket{0}$ and $\ket{1}$, which are identified with the standard (``computational'') basis of a 2-dimensional vector space.
Upon measurement in the computational basis $\{\ket{0},\ket{1}\}$, a qubit in the state $\ket{\psi} = \alpha\ket{0} + \beta\ket{1}$ collapses onto a classical outcome of
either $\ket{0}$ or $\ket{1}$, respectively, with probabilities $|\alpha|^2$ and $|\beta|^2$.
The magnitude of a quantum state vector is thereby always equal to one: $|\ket{\psi}|^2=|\alpha|^2+|\beta|^2=1$.
A state of many qubits can be represented by tensor products of single-qubit state vectors, as well as linear combinations of such tensor products.
Quantum operations are represented by matrices $\mathbf{U}$ that transform state vectors; these matrices are unitary, $\mathbf{U}\mathbf{U}^{-1}=\mathbf{U}^{-1}\mathbf{U}=1$.
A consequence of unitarity is that the number of input and output qubits to a quantum circuit are equal.

Single-qubit operations frequently used in quantum computation include the bit-flip operator $X$, which exchanges $\ket{0}$ with $\ket{1}$, the phase-flip operator $Z$, which takes $\alpha\ket{0}+\beta\ket{1}$ to $\alpha\ket{0}-\beta\ket{1}$, and the combined operation $Y=-i Z\cdot X$.
These primitives, along with the identity gate $I$, are referred to as Pauli gates.
When a Pauli operation has a control qubit, $CX$, $CY$, $CZ$, the two-qubit operation transforms the state of the target qubit with a Pauli gate depending on the state of the control.

\subsection{Quantum Applications}
\label{sec:q-apps}

 In the near term, one of the most promising applications for QCs are variational quantum algorithms (VQAs). In a VQA, a cost function is defined and encoded into a parameterized quantum circuit, referred to as an \emph{ansatz}. VQAs can be thought of as hybrid algorithms: a classical optimizer tunes the ansatz parameters based on circuit outcomes over many QC evaluations, working toward a solution, the quantum state prepared by the circuit, that either minimizes or maximizes the target cost function. VQAs have a wide range of applications including quantum chemistry~\cite{peruzzo2014variational} and optimization~\cite{moll2018quantum}. Two key type of VQAs are the variational quantum eigensolver (VQE) and the quantum approximate optimization algorithm (QAOA). These algorithms are well-poised for near-term demonstrations of quantum advantage because of their ability to adapt to the intrinsic noise profile of a QC.

Current quantum hardware is limited in that it (a) assigns each logical qubit of an algorithm to a single physical qubit in hardware, and (b) suffers from noise (i.e., hardware errors).
Indeed, many quantum algorithms, such as Shor's algorithm for quantum factoring~\cite{shor1999polynomial} and quantum algorithms that accelerate solving linear equations~\cite{harrow2009quantum}, are extremely sensitive to noise, which can irreversibly corrupt the result of a computation.
A central goal in quantum hardware design is therefore to achieve physical error rates below the thresholds required for fault-tolerant quantum error correction (QEC).
In fault-tolerant QEC, a logical qubit is encoded into the state of many physical qubits, in such a way that physical errors can be diagnosed and corrected without corrupting the state of the logical qubit.

Current QC error rates are rapidly approaching, and have in some cases surpassed, the theoretical thresholds required to run the surface code ($\sim 1\%$)~\cite{fowler2012surface}.
Going forward, achieving useful, fault-tolerant, error-corrected quantum computation requires further reducing physical error rates while simultaneously scaling up qubit numbers.

\subsection{Clifford Circuits}
\label{clifford-circuits}

The Clifford group is a set of quantum computing operations that transform Pauli strings (i.e., tensor products of single-qubit Pauli operators) into Pauli strings by conjugation; that is, if $P$ is the set of Pauli strings and $C$ the Clifford group, then $cpc^{-1}\in P$ for all $p\in P$ and $c\in C$~\cite{gottesman1998theory}.
The Clifford group consists of single-qubit Pauli gates, the square roots of these gates ($\sqrt{X}$, $\sqrt{Y}$, $\sqrt{Z}$), the Hadamard gate $H=(Z+X)/\sqrt{2}$, the two-qubit controlled-$X$ gate, and all operations that can be obtained by composing these gates.
Smaller qubit rotations that allow fine-grain control of qubit state, such as $Z^{1/4}=T$ are non-Clifford. The Clifford group does not provide a universal set of quantum gates and cannot be used for arbitrary quantum computation~\cite{gottesman1998heisenberg}. However, there are important quantum domains that have applications focused on the Clifford-space, including quantum networks~\cite{Veitch2014}, error-correcting codes~\cite{QECIntro}, teleportation~\cite{gottesman1999demonstrating} and error mitigation~\cite{czarnik2020error,strikis2021learningbased}.

A striking and important property of the Clifford group, summarized by the Gottesman-Knill theorem, is that quantum circuits consisting solely of Clifford operations are efficiently simulable on a classical computer~\cite{gottesman1998heisenberg}. 
This insight has been applied toward quantum circuit optimization such as dynamical decoupling scheduling~\cite{das2021adapt} and improved ansatz initialization for variational quantum algorithms (VQAs)~\cite{ravi2022cafqa}. While using only Clifford gates provides the advantage of tractable simulation, the use of a non-universal gate set during quantum computation prohibits the quantum system from exploring the full richness of the quantum state space.

\section{Prior Work and Motivation}
\label{sec:prior-work-mot}

In this section, we present state-of-art techniques for quantum circuit simulation with classical hardware. We also describe the theory that enables quantum circuit cutting. Both of these techniques provide inspiration for the SuperSim framework.

\subsection{Classical Simulation of Quantum Circuits}
\label{sec:classical-sim-quantum-circuits}

Quantum circuit simulation with classical hardware offers an effective means to quantify performance and troubleshoot potential issues within quantum algorithms without needing direct access to quantum hardware. Additionally, quantum circuit simulators that faithfully reflect real machine noise are a viable pathway to identify QC hardware features critical to the success of quantum algorithms. Unfortunately, state-of-art classical methods for quantum circuit simulation suffer from scaling challenges. As algorithms increase in qubit count $n$ (``circuit width''), the memory requirements to store a quantum state vector grow exponentially, $2^n$. Even using high-performance supercomputers, state vector simulation thus rapidly becomes intractable for growing qubit numbers. Current classical quantum circuit simulations appear to be bounded at $61$ qubits for quantum application circuits\cite{chongfsqc} and, conditionally, $100$ qubits for random quantum circuits\cite{liu2021}. In addition, noisy simulation that attempts to model the coherence and gate characteristics of real quantum hardware escalates the complexity of classical quantum circuit simulation even further.

Improving the viability of quantum circuit simulation on classical hardware is an active area of research. Many approaches have been proposed, such as those based on state vector representation~\cite{de2007massively}, Feynman path integrals~\cite{koh2017computing}, decision diagrams~\cite{grurl2022noise}, $ZX$ calculus~\cite{kissinger2022simulating}, and tensor networks~\cite{vidal2003efficient, markov2008simulating, schollwock2011density, pednault2017breaking, lykov2022tensor}. Exact quantum simulators, such as statevector simulators, have especially poor scaling because they faithfully capture quantum state in its entirety. Approximate simulators, like tensor network simulators based on matrix product states~\cite{zhou2020limits}, can exchange accuracy for scalability, producing outcomes with error margins.
Approximate methods can be exceedingly accurate in restricted scenarios, such as low or geometrically local entanglement, but these methods can fail catastrophically outside their limited regimes of applicability.

Prior work estimates that at minimum, millions of high-quality physical qubits will be required for fault-tolerant quantum computation~\cite{o2017quantum}. These qubits will implement error correcting codes, and since QEC circuits primarily consist of Clifford operations, classical simulation of $n$ qubits executing QEC code cycles is possible in time $poly(m,n)$, where $m$ is the number of gates in the circuit~\cite{kerzner2021clifford}. Quantum circuits including tens of thousands of qubits and millions of operations can be evaluated with Clifford simulation frameworks such as Stim~\cite{gidney2021stim}. In a Clifford simulator, the noise model can only include Pauli channels, which can be thought of as probabilistic Pauli operations interspersed throughout a circuit.
Clifford simulators are incapable of representing more realistic non-Clifford errors that occur in real quantum hardware.

\begin{figure}[t]
     \centering
         \includegraphics[width=0.8\linewidth,trim={0cm 0cm 1cm 1cm},clip]{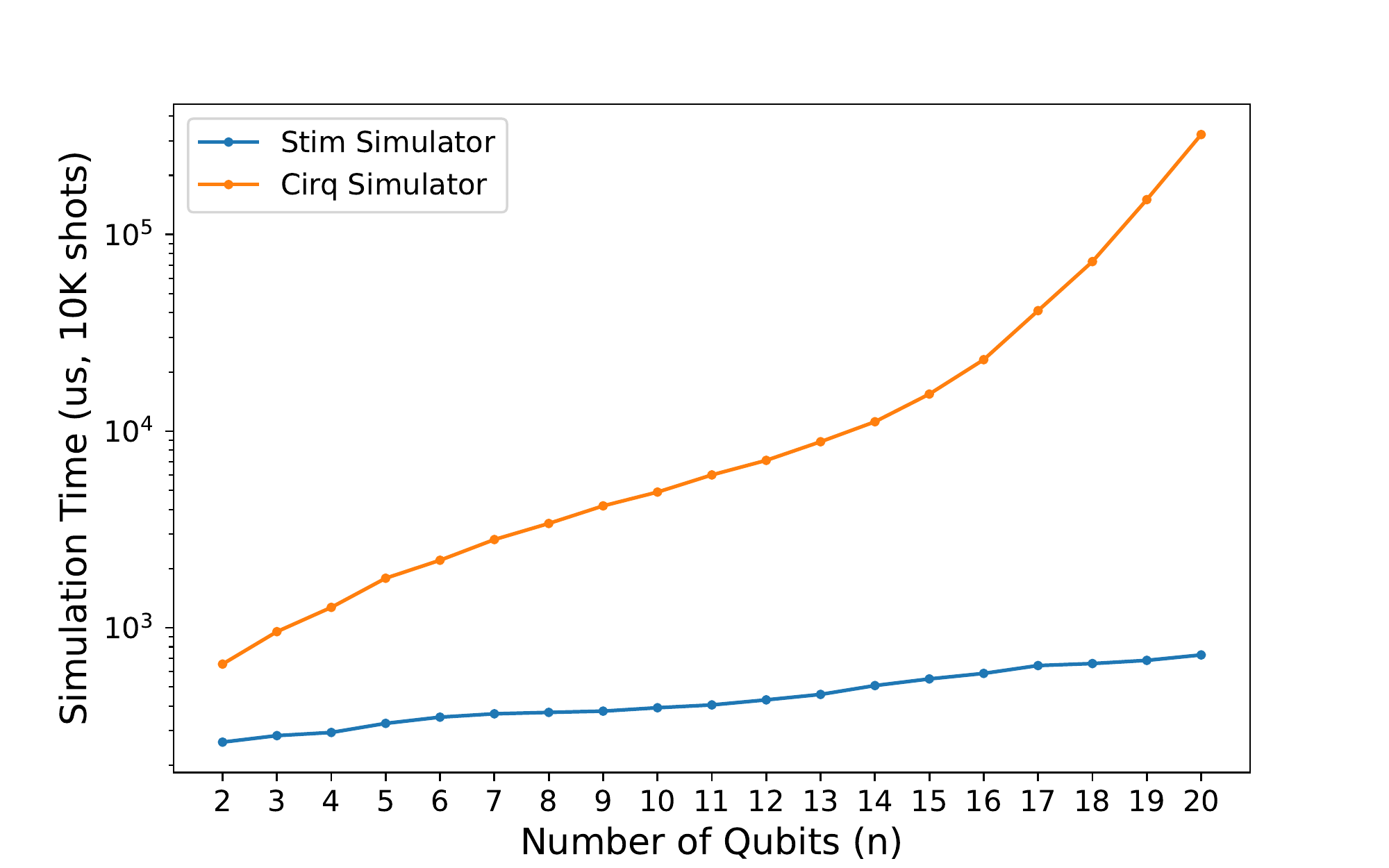}
        \caption{Simulation time vs. number of qubits for randomly generated Clifford circuits. Every circuit has a depth equal to width (number of qubits), and runtimes are averaged over 100 circuits.
        The Clifford circuit simulator runs significantly faster, showing more favorable scaling with qubit number.}
        \label{fig:stim-cirq-compare}
\end{figure}

Stim is a high performance simulation framework for quantum stabilizer circuits~\cite{gidney2021stim}. Fig.~\ref{fig:stim-cirq-compare} provides insight into the runtime differences between Clifford and (naive) non-Clifford simulation with randomly generated circuits that range from 2 to 20 qubits in size.
The depth of these circuits is equal to their width (qubit number), and results are averaged over 100 randomly generated circuits.
The Clifford circuit simulator, Stim, runs significantly faster than the statevector simulator. The significant performance improvements of the Stim simulator motivates its integration within SuperSim. However, additional methods are required to handle non-Clifford circuit elements during SuperSim simulation. 

When the number of qubits in a circuit is large but the number of non-Clifford operations comparably small, the classical simulation complexity of quantum circuits scales polynomially according to number of qubits and Clifford gates but exponentially with the addition of each non-Clifford gate. Thus, recent work has explored how to efficiently simulate near-Clifford circuits, or quantum circuits containing a few non-Clifford gates, typically chosen to be $T$ gates. Examples of Clifford+T simulators include techniques that use low-rank stabilizer decompositions~\cite{bravyi2019simulation} and Monte Carlo sampling of the stabilizer states produced by Clifford circuits~\cite{pashayan2022fast}. Note that Clifford+T provides a universal gate set for quantum computation, although it may require many $T$ gates to synthesize an arbitrary single qubit gate, such as $Z^a$ with $a<1/4$, to a high precision~\cite{dawson2005solovay}, significantly inflating the cost of simulation.

\subsection{Quantum Circuit Cutting}
\label{sec:circuit-cutting}

\begin{figure}[t]
\centering
\includegraphics[width=0.8\linewidth,trim={0cm 0cm 0cm 0cm},clip]{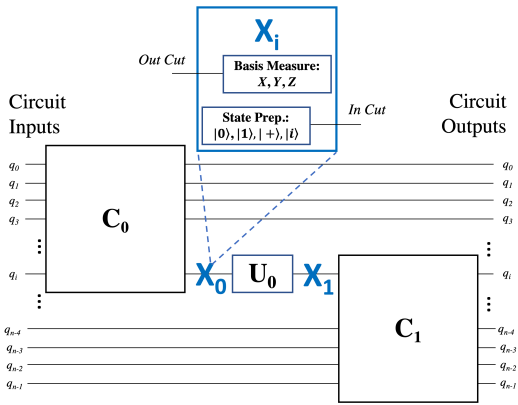}
\caption{Example of circuit cuts that isolate Clifford subcircuits, $\mathbf{C_i}$, from a non-Clifford operation, $\mathbf{U_0}$, in a near-Clifford circuit.
More generally, Clifford sub-circuits do not have to be cleanly separable: it is also possible to cut a non-Clifford gate from the middle of an otherwise Clifford circuit.
Such a cut would have the effect of increasing the size of a circuit fragment, but at the benefit of making the fragment efficiently simulable on a classical computer.
}
\label{fig:cutting-overview}
\end{figure}

Circuit cutting is a technique that enables dividing a large quantum circuit into subcircuits, or \textit{fragments}, that can be evaluated independently and recombined with classical postprocessing~\cite{peng2020simulating}.
The theory of circuit cutting relies on the ability to decompose arbitrary quantum operations into an orthonormal basis, such as the Pauli group.
The mechanics of circuit cutting involve identifying cuts that split a circuit into disjoint fragments, and evaluating the fragments either with a classical simulator or a QC.
An example of a quantum circuit cut into three fragments, two Clifford and one non-Clifford, is pictured in Fig.~\ref{fig:cutting-overview}.
If a cut appears at the end of a fragment, the corresponding qubit must be measured in various bases.
Similarly, if a cut appears at the beginning of a fragment, the corresponding qubit must be prepared in a variety of initial states.
As a consequence, each fragment has a number of \textit{variants}, where one variant corresponds to a fixed choice of initial states and measurement bases, that is exponentially large in the number of cuts that are incident to that fragment.
Additionally, postprocessing fragment execution data to reconstruct the output of a circuit has a computational cost that grows exponentially ($4^k$) with the number of cuts ($k$), making classical reconstruction of a circuit intractable for large numbers of cuts~\cite{perlin2021quantum}.
Note that in a Clifford+T circuit, the number of cuts required to remove all T gates is upper bounded by twice the number of T gates.

Computational overheads notwithstanding, past interest in circuit cutting has been driven by two primary motivations:
(a) theoretical interest in the trade-offs between quantum and classical computing costs, and
(b) reducing the size of circuits that are executed on quantum hardware, thereby allowing the execution of circuits that have more qubits than a given quantum computing device~\cite{perlin2021quantum, tang2021cutqc, uchehara2022rotation}.
Circuit cutting has also been shown to reduce the effects of statistical error that is intrinsic to quantum computing \cite{perlin2021quantum}. Finally, prior work has applied circuit cutting to error mitigation -- since simulation is noise-free, ideal outcomes of subcircuit simulation can be combined with noisy QC outcomes to form an error mitigated solution~\cite{liu2022classical}.

When a QC is targeted, many circuit cutting frameworks optimize processing by optimizing the location of cuts to minimize the overhead of classical fragment recombination.
We introduce a new application of circuit cutting, namely to the simulation of near-Clifford circuits.
Clifford circuits composed of thousands of qubits can be efficiently simulated on a classical computer~\cite{aaronson2004improved}, motivating SuperSim cut optimization that isolates non-Clifford gates. In a near-Clifford circuit, non-Clifford gates can be removed using a small number of cuts, and the resulting non-Clifford fragments can be small in terms of qubit and gate count, allowing them to be rapidly evaluated even on exact classical simulators.
Thus, an important challenge for our proposed Clifford-cut simulation framework is determining applications for which SuperSim provides a significant and useful advantage.

\section{Applications of Near-Clifford Simulation}
\label{sec:applications-of-near-clifford-simulation}

We developed SuperSim motivated by four near-Clifford circuit applications, which are described below.
Accordingly, the numerical results in Section \ref{sec:results} draw from benchmark circuits relevant to these applications.
Our development of SuperSim arose organically, as a result of four internal efforts that were all bottlenecked by the difficulty of simulating near-Clifford circuits. The first and foremost effort applies to simulations for quantum error correction (QEC), a task which has already inspired several years of prior research into near-Clifford simulation. In this context, SuperSim can enable researchers to more accurately characterize and diagnose QEC under realistic noise models. The other applications outlined below are different in nature and pertain to recently discovered end-user applications of near-Clifford quantum circuits. In these contexts, SuperSim will enable end users of quantum computation to benchmark the capabilities of quantum hardware.

\subsection{Simulations for Quantum Error Correction (QEC)}
The best known application of simulating near-Clifford circuits is toward estimating the performance of a QEC circuit against a non-Clifford noise model. In particular, in order to numerically validate a fault-tolerant QEC code, one must run repeated simulations of error correction circuits with stochastic errors injected according to a physical noise model.
Even with sophisticated tensor network methods, direct state vector or density matrix simulations have classical computing costs (as measured by both memory footprint and the number of floating-point operations) that scale exponentially with the number of qubits, precluding the numerical investigation of error-correcting codes that involve more than a few dozen qubits.

The dominant approach is therefore to perform stabilizer simulations. While these simulations are compatible with many typical error correcting codes (in that all gates are Clifford gates), the only errors that are supported in this model are Pauli errors, which are highly idealized and cannot capture many physically relevant sources of error in a quantum computer. For example, prevalent noise channels such as amplitude damping and overrotation cannot be captured by stabilizer simulations \cite{bennink2017unbiased}. Stabilizer-compatible approximations to these noise channels have been shown to substantially underestimate the impact of noise \cite{darmawan2017tensor}. For example, in \cite{darmawan2017tensor}, it was observed that for a width-5 surface code lattice modeled with a systematic $1^\circ$ overrotation noise, the estimated logical error rate is $10^{-16}$ based on a Pauli noise approximation, whereas the actual logical error rate is $10^{-6}$ -- ten orders of magnitude greater. 

Simulation approaches for QEC are required that can scale to hundreds of qubits without needing an approximated variant of the noise. In the case of Clifford QEC circuits, even adding a few non-Clifford gates to create a near-Clifford QEC description could improve modeling, creating a more granular picture of the impact error has on quantum computation. A primary motivation for the development of SuperSim was delivering a near-Clifford quantum circuit simulator that helps fill the gap that exists between approximate and realistic noisy QEC simulation.
 
\subsection{Variational Optimization with CAFQA}
\label{sec:variational optimization}

 As introduced in Section~\ref{sec:q-apps}, VQAs are hybrid quantum-classical algorithms in which a classical optimization subroutine makes queries to a QC in the form of gate parameters. The information the QC sends as a response then helps inform the classical optimizer's next quantum query. An ansatz is the parameterized quantum circuit used within the aforementioned feedback loop, and there are many ansatz structures that can be used within a VQA. 
 
VQE is a type of VQA often applied in quantum chemistry applications. There are a variety of different types of circuit structures for VQE, and the hardware efficient ansatz (HWEA) is particularly promising as it is customizable to a targeted QC, making it low depth and well-suited for near-term quantum hardware~\cite{kandala2017hardware}. At its core, the HWEA consists of single-qubit rotation gates (the rotation angles are the tunable parameters), a layer of two-qubit entangling gates, and a final layer of single-qubit rotation gates. Alternatively, QAOA is often applied toward optimization problems, such as MaxCut. Here, we study the Sherrington-Kirkpatrick (SK) model for MaxCut
on complete graphs with edge weights randomly drawn from
$\{-1, +1\}$~\cite{farhi2022quantum}. Our QAOA ansatz matches the SK model exactly, meaning that it requires all-to-all connectivity in the quantum circuit structure.
 
 The expressiveness of a QC's gate set allows VQAs to be more precise. In addition, more layers in the ansatz (i.e., single-qubit rotations, entangling operations, and single-qubit rotations) help boost VQA accuracy. A challenging problem for both VQE and QAOA is how to initialize the parameters at the start of VQA execution. Parameters can be randomly initialized, but finding approximately optimal angles for the single-qubit gates can help the VQA more rapidly and accurately converge on a desired answer. As a solution, the work in~\cite{ravi2022cafqa} presents the Clifford Ansatz for Quantum Accuracy (CAFQA) that uses an initial ansatz containing only Clifford operations, optimizing ansatz parameters with efficient Clifford simulation. This Clifford-based optimization for VQA demonstrated promising results - convergence to the desired VQA solution was found to be quick and accurate. However,~\cite{ravi2022cafqa} recognizes that the addition of just a single $T$ gate allows the ansatz tuning space to become richer, opening doors for greater VQA accuracy. As discussed and preliminarily evaluated in~\cite{ravi2022cafqa}, more optimal VQA initialization results if a few non-Clifford gates are permitted in the VQA ansatz initialization via classical simulation. Unfortunately, the addition of a $T$ gate to a Clifford circuit removes opportunities for tractable scaling, motivating the search for practical Clifford+T simulation solutions. With SuperSim, new opportunities exist for near-Clifford variational optimization with a near-CAFQA ansatz initialization framework.

\subsection{Generative Modeling}
In \cite{anschuetz2022interpretable} and \cite{gao2022enhancing}, it was rigorously proven that Clifford circuits exhibit unconditional quantum advantages for generative modeling machine learning tasks. Typically, this advantage is quadratic, which coincides with the fact that classically simulating an $n$-qubit Clifford circuit requires tracking $O(n^2)$ stabilizer bits. Interestingly however, this quadratic advantage in memory can sometimes be lifted to a quartic $O(n^4)$ advantage in time \cite{anschuetz2022interpretable}, which has motivated deeper study of Clifford-based quantum speedups.

However, a challenge in both \cite{anschuetz2022interpretable} and \cite{gao2022enhancing} is to train these machine learning models, which requires non-Clifford gates. SuperSim is ideally matched to this context of primarily Clifford gates, with non-Clifford gates to enable gradient descent in the parameter landscape. We have initiated collaboration with the authors of \cite{anschuetz2022interpretable} and \cite{gao2022enhancing} to validate this approach.

\subsection{Fingerprinting}
Finally, we envision applying SuperSim to SupercheQ \cite{gokhale2022supercheq}, a protocol for quantum speedup in \textit{fingerprinting} that we released last year. The task of fingerprinting involves validating that two files are equal, against a worst-case adversarial model (which obviates even cryptographic hashes to avoid hash collisions). With just Clifford circuits, SupercheQ’s Incremental Encoding (IE) variant asymptomatically matches the best possible classical protocol in space, while attaining a quantum advantage in terms of \textit{incrementality}. We also show that SupercheQ’s Efficient Encoding (EE) variant, which is entirely non-Clifford, attains an exponential advantage over the best possible classical protocol, though it lacks incrementality. We thus propose using SuperSim to investigate the middle-ground, in particular by enriching the Clifford-only space of SupercheQ-EE with a few non-Clifford gates. Work towards studying this is under way by other members of the Super.tech team.

\section{SuperSim: A Circuit-cutting Quantum Simulation Framework}
\label{sec:SuperSim-framework-overview}

Motivated by the need for scalable quantum circuit simulation solutions and inspired by the performance of Clifford simulation, we developed the SuperSim simulation framework based on Clifford-based circuit cutting. SuperSim is part of the Super.tech family of quantum software solutions, nicely complimenting the SuperstaQ optimized ``write once, run anywhere'' compiler~\cite{superstaq} and the SupermarQ suite of application-centric benchmarks~\cite{tomesh2022supermarq}. SuperSim is on track to be an open-source and consumer ready product with a launch planned for summer 2023. The launch will encompass the release of the open-source codebase along with accompanying documentation, tutorials, and product framework that are currently under development.

The SuperSim framework is written in Python. Its three critical components include the circuit cutting algorithm, the fragment evaluator, and the probability distribution reconstruction algorithm. SuperSim uses Cirq~\cite{cirq_developers_2022_7465577} for representing and manipulating quantum circuits. The Clifford circuit evaluation invokes the Stim simulation backend while non-Clifford fragments use the Qsim statevector simulator by default~\cite{quantum_ai_team_and_collaborators_2020_4023103}, although the Cirq statevector simulator is also supported. In the rest of this section, the three main elements of SuperSim will be described.

\subsection{SuperSim Circuit Cutter}
The circuit cutter begins the SuperSim flow -- it is the first transformer that an input quantum circuit encounters. The cutter is fundamentally a circuit parser that identifies the Clifford and non-Clifford elements within a circuit. As input circuits are intended to be near-Clifford, locations for circuit cuts isolate the non-Clifford operations. After developing a set of cut locations, the original input circuit is separated into disjoint subcircuits, which are referred to as circuit fragments. The circuit fragments are then sent to the SuperSim fragment evaluation module.

\subsection{SuperSim Fragment Evaluator}
All circuit fragments are tagged as either Clifford or non-Clifford.
Clifford fragments are then passed to a Clifford circuit simulator (Stim), and non-Clifford fragments are passed to an exact (e.g.~statevector) simulator.
Each fragment induces several \textit{fragment variants}, where one variant of a fragment corresponds to the fragment with some additional operations attached at the locations of cuts incident to that fragment.
Fragment inputs and outputs are designated as follows: 
\begin{itemize}
    \item Circuit Input: An input of the original, uncut circuit. All circuit input qubits are initialized in the $\ket{0}$ state, so no additional operations are associated with circuit inputs.
    \item Circuit Output: An output of the original, uncut circuit. All circuit output qubits are measured in the computational basis $\{\ket{0}, \ket{1}\}$, so no additional operations are required prior to measuring the circuit outputs on fragments.
    \item Quantum Input: An input of the fragment, but not of the original circuit. Every quantum input is \textit{downstream} of a cut in the original circuit.  Additional operations are required to prepare different states at quantum inputs (see Fig.~\ref{fig:cutting-overview}).
    \item Quantum Output: An output of the fragment, but not of the original circuit. Every quantum output is \textit{upstream} of a cut in the original circuit.  Additional operations are required to measure quantum output qubits in different bases (see Fig.~\ref{fig:cutting-overview}).
\end{itemize}
After generating a batch of variants for each Clifford fragment, where one variant corresponds to a fixed choice of initial states and final measurement bases for that fragment, a simulator determines the probability over measurement outcomes at the circuit outputs of each variant.
These probability distributions, tagged by associated choices of initial states and measurement bases, are later classically postprocessed to reconstruct a probability distribution over measurement outcomes for the original circuit \cite{perlin2021quantum}.
While the current version of SuperSim simulates fragment variants sequentially, future iterations of SuperSim will support parallel simulation strategies that will yield significant reductions in runtime \cite{perlin2019parallelizing}.

In principle, tensor network methods can be used to collect all necessary non-Clifford fragment data directly without needing to simulate different fragment variants.
In practice, the non-Clifford fragments in the applications considered in this work are so small that this optimization would at best result in negligible performance improvements.
Nonetheless, to accommodate unforeseen applications for which a significant fraction of classical computing resources is devoted to non-Clifford fragment simulation, future iterations of SuperSim will use a tensor network simulator for non-Clifford fragments.

\subsection{SuperSim Distribution Builder}
The final key component of SuperSim combines fragment simulation data to build, or reconstruct, the probability distribution outcomes of the original, uncut circuit using the measurement data of the fragment variants.
The theoretical details of fragment data postprocessing and combination are provided in~\cite{perlin2021quantum}.
In a nutshell, postprocessing fragment data first involves applying maximum-likelihood corrections to build fragment models that are self-consistent, thereby mitigating the effect of sampling error that is unavoidable in probabilistic Clifford simulation methods, as well as circuit executions on real quantum hardware.
After constructing self-consistent fragment models, individual probabilities for measurement outcomes in the original circuit can be expressed as the result of a tensor network contraction, with one tensor per fragment.
This tensor network has a number of edges equal to the number of cuts in the original circuit, resulting in an overall computational cost to contraction that is roughly exponential in the number of cuts.
Altogether, the current version of SuperSim uses the codes provided in~\cite{perlin2021quantum} for postprocessing fragment data and building a probability distribution over measurement outcomes at the end of a recombined circuit.
Future iterations of SuperSim will leverage additional techniques to parallelize the post-processing of fragment data \cite{perlin2019parallelizing}, which is currently implemented in a sequential fashion.

As a final point, we emphasize that SuperSim does not rely on any approximations; its only source of inaccuracy is statistical error from sampling the outputs of Clifford circuit fragments. SuperSim is thereby guaranteed to be at least as accurate as the evaluation of a circuit on real quantum hardware, with which sampling error is fundamental and unavoidable. In addition, the SuperSim framework is readily amenable to so-called ``strong simulation'' of a quantum circuit, whereby the probability to observe a particular bitstring at the output of the circuit can be computed to machine precision without added computational overheads.

\section{Methods}
\label{sec:methods}

In this section, we describe our experimental environment used to analyze SuperSim performance. This includes the simulators compared against SuperSim, the chosen benchmark circuits, and the experimental platform.

\subsection{Included Simulators}

Performance of the SuperSim framework is compared to the simulators available in Qiskit~\cite{Qiskit}, IBM's open-source quantum software development kit, as well as to statevector simulation. As a note, the SuperSim framework builds a probability distribution from the fragment simulation results. As a result, we used all of the below simulators as samplers, using 5000 shots to build output distributions. 


\textbf{Qiskit Extended Stabilizer Simulator:} A Clifford+T simulator that is an implementation of~\cite{bravyi2019simulation}. The extended stabilizer uses ranked-stabilizer decomposition and is approximate. Complexity scales with the number of non-Clifford gates as non-Cliffords set the number of stabilizer terms. The $T$ gate (and any $Z^a$ operation) is included in the basis gate set. Up to 63 qubits are supported in a circuit.

\textbf{Qiskit Matrix Product State Simulator:} A tensor-network simulator that relies on Matrix Product States (MPSs). Tensor-network simulators in general are able to scale to larger quantum systems as compared to statevector simulators at the cost of depth. Thus this simulator can support more qubits than statevector and extended stabilizer simulators as long as entanglement between qubits is low and critical path is small. A large gate set is supported.

\textbf{Statevector (SV) Simulator:} statevector simulation, such as with the Qsim simulator, uses exact methods to represent all $2^n$ state amplitudes.

\subsection{Benchmark Circuits}

Simulator performance is currently evaluated with three benchmark circuit types, with varying numbers of qubits. These circuits were chosen according to the applications proposed in~\ref{sec:applications-of-near-clifford-simulation}.
In the future, we will sharpen these benchmarks, and additionally benchmark SuperSim against other simulators in a wider array of test circuits, as discussed in Section~\ref{sec:ongoing-roadmap-for-future-eval}.


\textbf{Near-Clifford VQE HWEA:} As discussed in Section~\ref{sec:variational optimization}, the near-Clifford HWEA is a helpful tool for VQE optimization. The base structure for this quantum circuit (or a single round) is a Clifford HWEA consisting of a single layer of single-qubit gates, a layer of entangling stages, and final layer of single-qubit gates.
Together, these three layers of gates constitute a single layer of the HWEA.
Circuit width (number of qubits), depth (total number of layers), and non-Clifford gate injections are varied during experimentation.

\textbf{Near-Clifford MaxCut QAOA:} Similar to the near-Clifford VQE HWEA, this benchmark is used for optimization of QAOA initalization. Since the ansatz matches the SK model exactly, each round requires a significant amount of qubit-qubit connectivity.

\textbf{Phase Flip Repetition Code:} The repetition code is a classical error correcting code that constitutes an important precursor to QEC. In our preliminary experiments to evaluate SuperSim performance for QEC applicaions, we use a single round of the phase code implementation included in SupermarQ \cite{tomesh2022supermarq}.

\subsection{Experiment Details}

SuperSim will be a product targeted for a wide range of end users, from students in an academic setting to users in industry. As such, results of this work were generated using both a MacBook Pro with a 2.4 GHz i9 Intel Processor (32 GB of RAM) as well as a Linux server with up to 16 vCPU (32 GB of RAM). 

To quantify accuracy, we frequently include simulation fidelity along with runtime. On dense distributions, like those often seen in VQAs, we use the Hellinger fidelity of marginal probability distributions for measurement outcomes on individual qubits, as compared to ideal statevector simulations. This is motivated by emerging research that supports training VQAs on a sum of local cost functions instead of a single global cost function for scalability~\cite{cerezo2021cost}. On sparse distributions (i.e. ones with one or few measured observables), we use Hellinger fidelity on the complete distribution.


\section{Results}
\label{sec:results}

\subsection{VQE Application Evaluation }

\begin{figure*}[t]
\centering
\includegraphics[width=0.8\linewidth,trim={2cm 0cm 1cm 1.75cm},clip]{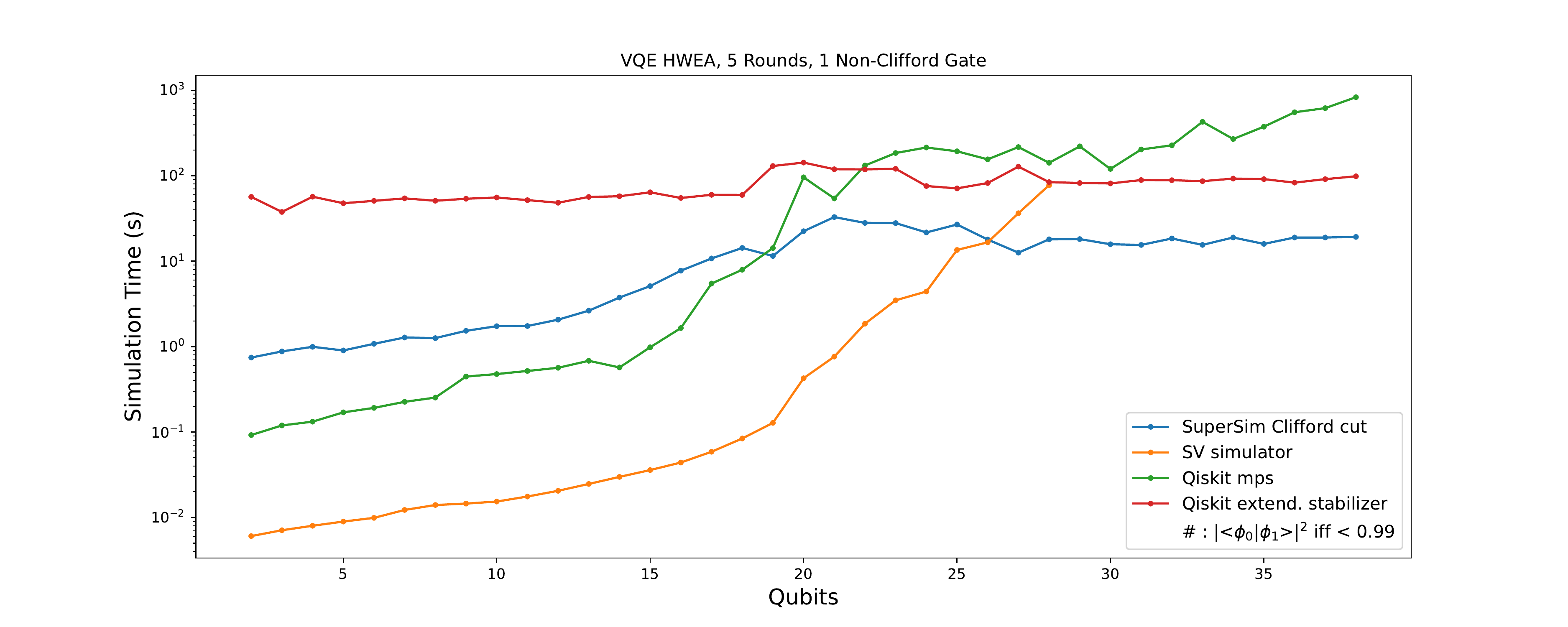}
\caption{Averaged simulation time as a function of qubit number for the VQE HWEA with 5 rounds and one randomly injected $T$ gate for a variety of simulators specified in the legend. Experiments were set to timeout at 30 minutes thus the SV simulator curve is truncated at 28 qubits.
SuperSim outperforms all other simulators past 26 qubits, even as SuperSim continues to have modest runtimes with hundreds of qubits (see Figure \ref{fig:supersim-HWEA-scaling}).}
\label{fig:vqa-analysis-time-vs-qubits-all}
\end{figure*}

\begin{figure}[t]
\centering
\includegraphics[width=0.8\columnwidth,trim={0cm 0cm 0cm 1.8cm},clip]{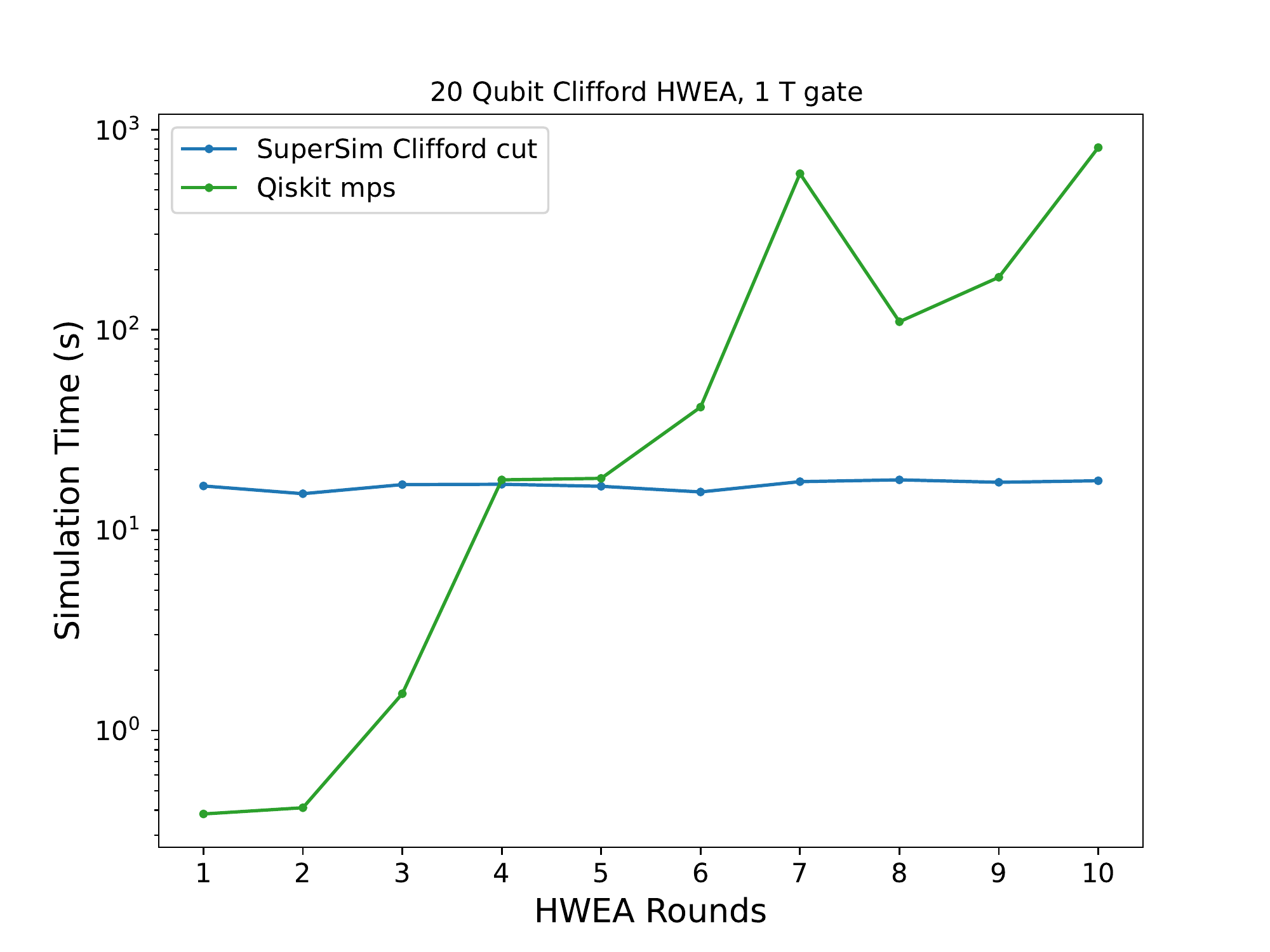}
\caption{Simulation time as a function of VQE HWEA rounds for 20 qubit HWEA base circuit. In every case, a single $T$ gate is injected into the circuit at a random location. This plot demonstrates SuperSim Clifford cut scalability with circuit depth and entanglement growth, in contrast to the Qiskit MPS Simulator.
Note that SuperSim simulation time in this experiment is insensitive to HWEA rounds because it spends the bulk of its time postprocessing fragment simulation data, rather than simulating fragments themselves.}
\label{fig:supersim-vs-mps-HWEA-depth-scaling}
\end{figure}

\begin{figure}[t]
\centering
\includegraphics[width=0.8\columnwidth,trim={0cm 0cm 0cm 1.8cm},clip]{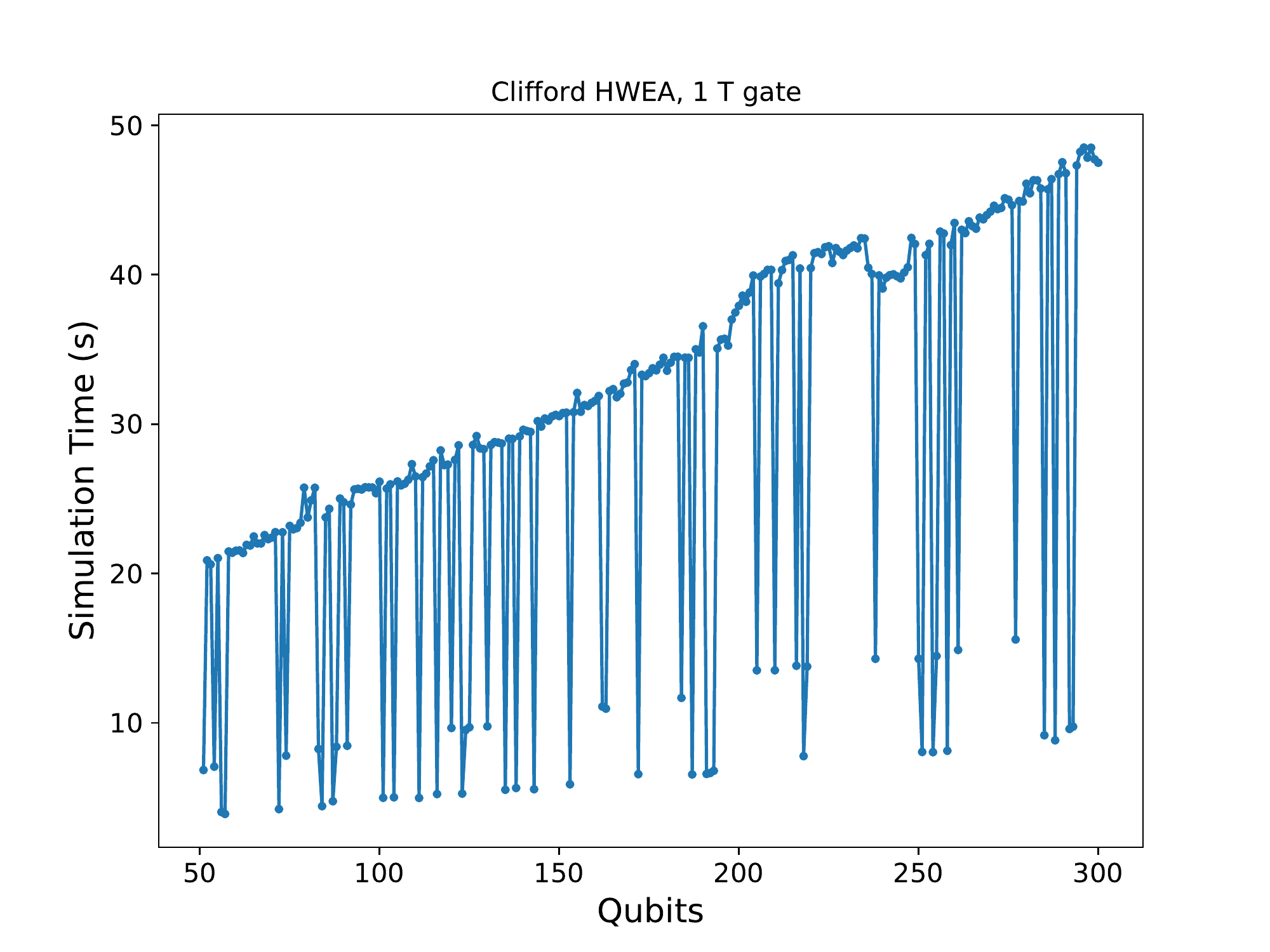}
\caption{Simulation time as a function of qubit number for the VQE HWEA with 5 rounds and one randomly injected $T$ gate.
These results are essentially the same experiment as in Figure \ref{fig:vqa-analysis-time-vs-qubits-all}, but for qubit numbers up to 300, demonstrating that that SuperSim can simulate circuits that are much larger than is possible with statevector and extended stabilizer simulators.
The ``noisy'' dependence of runtime on qubit number in this figure is an artifact of simulating, for each qubit number, only a single circuit with one randomly injected $T$ gate.
Generally speaking, $T$ gate location can significantly affect runtime for SuperSim simulation of near-Clifford circuits, especially as the location of the $T$ gate changes the number of fragments that need to be simulated and recombined.}
\label{fig:supersim-HWEA-scaling}
\end{figure}

Classical simulation of a near-Clifford HWEA can be applied toward the near-CAFQA framework proposed in Section~\ref{sec:variational optimization}.  To begin, we analyze the relationships between simulation runtime and circuit width. We generate VQE HWEA circuits that range from two to 38 qubits in size. Each circuit has five layers of entangling operations sandwiched between single-qubit parameterized (Clifford) operations. The results for time vs. qubits for the HWEA benchmark with five HWEA layers and one randomly injected $T$ gate are provided in Fig.~\ref{fig:vqa-analysis-time-vs-qubits-all}. Results average multiple simulation experiments (five) for a generated benchmark and randomly placed $T$ gate. The SuperSim Clifford cut simulator, Qiskit extended stabilizer simulator, and Qiskit MPS simulator are compared to statevector simulation. Accuracy is included in this analysis - all data points had an fidelity higher than $\sim 0.99$ when compared to statevector outcomes.
For smaller circuits, the runtime of SuperSim is larger than all of the simulators by nearly an order of magnitude, with the exception of qiskit extended stabilizer. However, at about 25 qubits, we observe a crossover where SuperSim becomes the most efficient simulator in terms of execution time.
These results were generated using server-based resources.

In the next analysis based on the near-Clifford VQE HWEA benchmark, we focus on SuperSim and MPS simulator performance relationship with circuit depth. To this end, Fig.~\ref{fig:supersim-vs-mps-HWEA-depth-scaling} shows SuperSim and MPS runtime scaling as the near-Clifford HWEA with a fixed width of 20 qubits, but a depth that varies from one to 10 layers of the HWEA, in every case with one injected $T$ gate. We observe that SuperSim begins to demonstrate a strong advantage over MPS at around 6 layers of the HWEA. These results were generated using a laptop computer.

For our third experiment with the near-Clifford VQE HWEA benchmark, we study large-scale SuperSim scaling - up to 300 qubits in system size. Fig.~\ref{fig:supersim-HWEA-scaling} shows the scalability of the SuperSim framework based on Clifford circuit cutting. Results for simulation time as a function of qubit number for HWEA with 5 rounds, up to 300 qubits, and one randomly injected $T$ gate. Fig.~\ref{fig:supersim-HWEA-scaling} was produced with a laptop, and these results show that SuperSim can simulate circuits that are much larger than are feasible with statevector and extended stabilizer simulators. Furthermore, the runtimes included in Fig.~\ref{fig:supersim-HWEA-scaling} are generally smaller than the runtimes of other Fig.~\ref{fig:supersim-HWEA-scaling} methods past 30 qubits. We note that the runtimes in Fig.~\ref{fig:supersim-HWEA-scaling} do not increase monotonically with qubit number.
This behavior is an artifact of simulating, for each data point in Fig.~\ref{fig:supersim-HWEA-scaling} represents only a single circuit with one randomly injected $T$ gate.
If a $T$ gate happens to cleanly split a circuit into two Clifford sub-circuits, for example, the resulting SuperSim simulation takes considerably less time than if a $T$ gate is cut out of the middle of a single large Clifford circuit.

\begin{figure*}[t]
\centering
\includegraphics[width=0.8\linewidth,trim={2cm 0cm 1cm 1.75cm},clip]{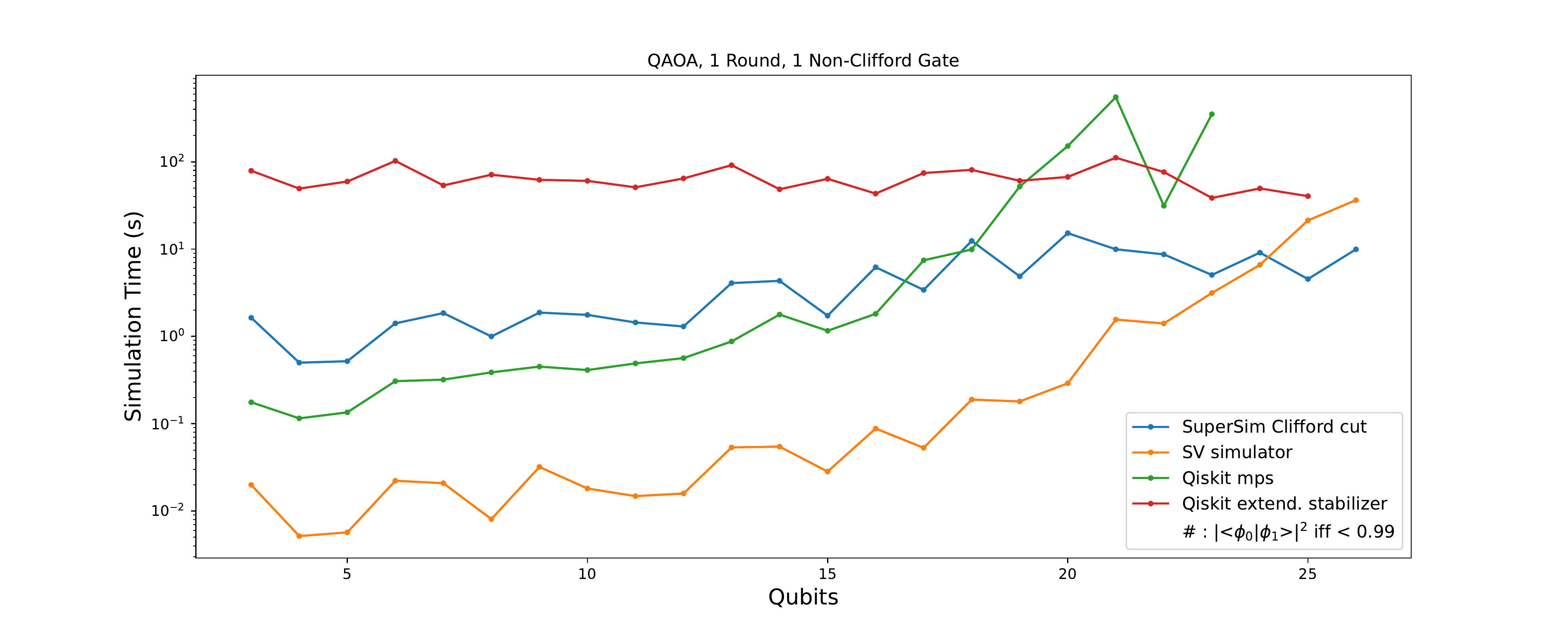}
\caption{Averaged simulation time as a function of qubit number for QAOA MaxCut with one round and one randomly injected $T$ gate for a variety of simulators specified in the legend. Experiments were set to time out at 30 minutes.}
\label{fig:qaoa}
\end{figure*}

\subsection{QAOA Application Evaluation}

In this section, we analyze SuperSim's performance on the near-Clifford MaxCut QAOA benchmark. This evaluation, seen in Fig.~\ref{fig:qaoa}, was completed on server-based compute resources, and experiment timeout was set to 30 minutes. Once again, we analyze the relationship between simulation runtime and circuit width. Accuracy is included in this analysis - all data points had an fidelity higher than $\sim 0.99$.

The QAOA benchmark is generated with sizes ranging from three to 26 qubits. One round of QAOA is implemented that features all-to-all connectivity among circuit qubits, and one $T$ gate is randomly injected into the circuit. To generate each datapoint in Fig.~\ref{fig:qaoa}, five experiments are executed, and results are averaged to create each runtime datapoint / fidelity outcome. Similar to the results in Fig.~\ref{fig:supersim-HWEA-scaling}, we initally see MPS and the statevector simulator outperforming SuperSim, but a point of crossover is eventually reached. These results further reinforce the viability of SuperSim in VQA optimization.

\subsection{QEC Application Evaluation}

\begin{figure*}[t]
\centering
\includegraphics[width=0.8\linewidth,trim={2cm 0cm 1cm 1.75cm},clip]{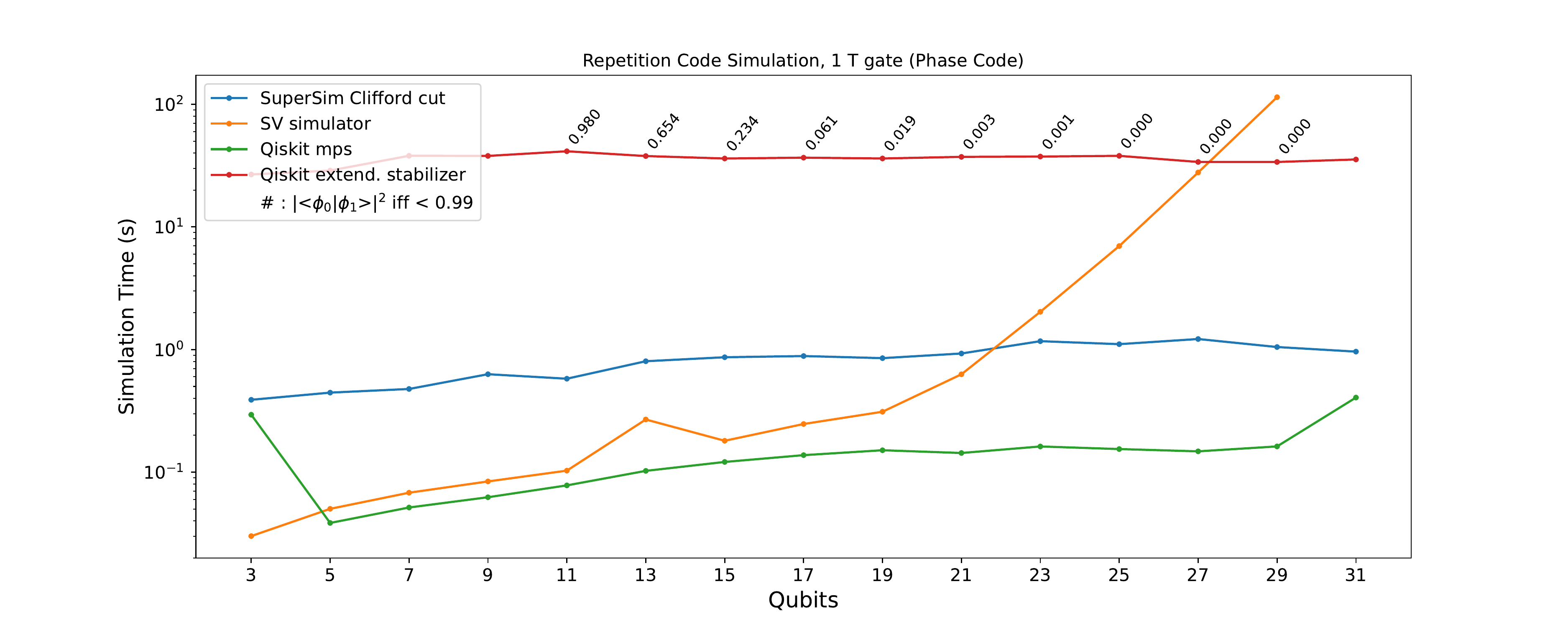}
\caption{Simulation time as a function of qubit number in the simulation of a single phase-correcting repetition code cycle for a variety of simulators specified in the figure legend. Points are annotated where the fidelity of simulation results compared to exact statevector simulations are at or below 0.99.
Note that the MPS simulator outperforms SuperSim in this benchmark due to the fact that repetition code circuit generates very little entanglement between qubits.
Future benchmarks of SuperSim will include fullly quantum error correcting codes in high entanglement regimes that are known to be out of reach for MPS \cite{vidal2003efficient, cirac2021matrix}.}
\label{fig:rep-code}
\end{figure*}

This section describes preliminary results for SuperSim performance in the domain of QEC.
Here, we use the SupermarQ implementation of phase repitition code and study runtime scaling with increasing qubit number. 

We note that our use of the phase code is an artificial ``proxy'' benchmark that provides merely preliminary results for QEC circuit simulation - this benchmark only addresses phase flip errors, not bit flip errors, and it merely detects errors without correcting them. Fig~\ref{fig:rep-code} shows runtime as a funciton of qubit number for one round of the phase repetition code with one randomly injected $T$ gate. On the plot, points are annotated where fidelity to statevector simulation results are less than 1.00. In other words, if a point on the curve does not have a label (as seen in Figs.~\ref{fig:supersim-HWEA-scaling},~\ref{fig:qaoa}), its fidelity is greater than 0.99. In this experiment, a significant disparity was noted in the accuracy of the extended stabilizer simulator. 

This initial step toward QEC provides insight toward the benefits of SuperSim: our framework scales favorably with qubit number, while most other simulators have poor runtime scaling, taking longer than SuperSim past 25 qubits.
Moreover the extended stabilizer simulator quickly becomes useless in terms of accuracy for the phase-repetition code.

A clear exception to the favorability of SuperSim over other methods in Figure \ref{fig:rep-code} is the MPS simulator, which outperforms SuperSim for all qubit numbers provided.
This exception is an artifact of the fact that the circuit for a phase repetition code cycle generates very little entanglement between the qubits involved.
Whereas SuperSim performance is unaffected by the degree of entanglement generated in a Clifford circuit, MPS is well-known to fail catastrophically in terms of runtime, memory footprint, accuracy, or a combination of all three (depending on the precise implementation of MPS) in the regime of high and nonlocal (``volume-law'') entanglement \cite{vidal2003efficient, cirac2021matrix}.
Near-future benchmarks that are more faithful to QEC simulation performance will involve codes that also correct bit-flip errors, thereby requiring the simulation of highly entangled states that are simply out of reach for MPS.

\section{Discussion}
\label{sec:discussion}

The alpha version of the SuperSim framework was motivated, presented, and evaluated in this paper. The novelty of this framework is that it combines the benefits of quantum circuit cutting and Clifford quantum circuit simulation while knowing that both of these techniques suffer from their unique constraints. Circuit cutting scales poorly, and we note that the complexity of SuperSim is bounded by twice the number of non-Clifford gates, having an overall simulation cost that is exponential in the number of non-Cliffords. Further, the Clifford-only application domain is limited. Nonetheless, we identify near-Clifford quantum circuits as a target for accelerated simulation that balances tradeoffs with complexity. As a note, although our evaluation focused on injecting $T$ ($=Z^{1/4}$) gates into benchmarks, the SuperSim circuit cutter is able to identify, isolate, and create fragments for other non-Clifford gates, such as arbitrary $Z^a$ operations. 

The benefits of quantum circuit simulation via SuperSim have potential to unlock improvements in the debugging, performance quantification, and noise tolerance evaluation of quantum algorithms. Further, SuperSim will help guide hardware development by pushing the limits of circuit simulation, which is essential for co-design, assessing architectural choices in hardware, and pinpointing hardware features that are critical to the success of algorithms. We have ongoing efforts to optimize and improve the SuperSim framework, but it already shows potential as a domain-specific quantum circuit simulator. Our initial evaluation of SuperSim targeted VQA optimization and QEC applications, and our results highlight instances where SuperSim shows advantage over alternative open source quantum circuit simulation solutions. For instance, the experiments in Section~\ref{sec:results} show that compared to the statevector simulation, SuperSim demonstrates significantly lower runtimes as the VQE HWEA, QAOA MaxCut, and repetition code benchmarks approach 30 qubits in size. Based on the discussion of Section~\ref{sec:classical-sim-quantum-circuits}, this is an expected result. However, in both the Fig.~\ref{fig:supersim-HWEA-scaling} and Fig.~\ref{fig:rep-code} experiments, SuperSim outperforms the Qiskit extended stabilizer simulator, which is significant as this simulator is based on state-of-art Clifford+T simulation methods~\cite{bravyi2019simulation}. We highlight that in some cases, SuperSim demonstrates runtime improvments of around $100\times$ as compared to the Qiskit extended stabilizer. Further, the results in Fig.~\ref{fig:rep-code} show that SuperSim outshines Qiskit extended stabilizer in accuracy.

SuperSim can scale to hundreds of qubits, Fig.~\ref{fig:supersim-HWEA-scaling}, and simulate these large quantum circuits in tens of seconds.
We must highlight, though, that our benchmarks also have large-qubit regimes (Fig.~\ref{fig:rep-code}) in which the Qiskit MPS simulator outperforms SuperSim.
This finding is not surprising given the nature of benchmarks chosen, as MPS is an approximate simulation scheme that is well-known to perform well in (and was, in fact, designed for) low-entanglement regimes, while failing to provide accurate results for the simulation of highly entangled states \cite{vidal2003efficient, cirac2021matrix}.
This fact is seen clearly as the HWEA circuit increases in depth in Fig.~\ref{fig:supersim-vs-mps-HWEA-depth-scaling} and MPS simulation time grows exponentially.
In future work, we will investigate more near-Clifford quantum circuits that are highly entangled, including fully quantum error correcting circuits (as opposed to classical repetition codes), as candidate SuperSim applications. We also plan to explore relationships between number of non-Clifford gates within a near-Clifford circuit and simulation runtime. 



\section{Ongoing: Clifford-specific Cutting Optimizations} \label{sec:clifford_specific_cutting_optimizations}
While the circuit cutting procedure described so far effectively leverages fast Clifford simulation at a per-subcrircuit level, the procedure for stitching together results has thus far been agnostic to whether subcircuits are Clifford or not. As it turns out, breaking this abstraction barrier leads to substantial performance improvements. As a concrete example, we consider a Clifford subcircuit $C_i$ upon which we must measure a cut qubit in multiple bases—for instance, $C_0$ in Fig.~\ref{fig:cutting-overview}. Following the standard circuit cutting procedure, we must make Pauli observable measurements $\braket{P}$ after running $C_i$. We next describe two Clifford-specific cutting optimizations: (1) significantly fewer requisite shots and (2) fewer downstream stitching calculations.

The first optimization—fewer requisite shots—stems from the observation that for Clifford circuits, $\braket{P}$ is either $-1$, $0$, or $+1$ \cite{mike_ike_2020}. After rotating to computational basis measurements, these correspond to being in the $\ket{1}$, $\ket{+}$ (or any other equal superposition), and $\ket{0}$ states respectively. It turns out this significantly reduces the number of shots that are needed to determine $\braket{P}$ versus generic estimation where accuracy would scale with the square root of number of shots. By analogy, our setting is akin to being promised that we have a completely-biased heads coin (-1), fair coin (0), or completely-biased tails coin (+1). If we flip the coin a few times and find at least one heads and at least one tails, we immediately know we have a fair coin. In the other two cases, we will get only heads or only tails. In a similar fashion, in our setting, if we measured just 10 shots but found $\ket{1}$ ($\ket{0}$) every time, we can be quite confident that $\braket{P}$ is -1 (+1).

The second optimization—fewer downstream stitching calculations—stems from the observations that in Clifford circuits, the output state must have $\braket{P} = 0$ for many Pauli observables. Consider the one-qubit case, where the six Clifford states belong on the Bloch octahedron. For each of these states, two of $\braket{X}$, $\braket{Y}$, and $\braket{Z}$ must be equal to 0. Plugging in zeros to the circuit cutting stitching equations, we find that some of the ``downstream’’ circuits that would otherwise need to be evaluated can be skipped. Moreover this phenomenon scales favorably: for multi-qubit observables, the fraction of non-zero Pauli observables approaches 0, significantly reducing downstream calculations.

Pending the careful theoretical development and software implementation of these ideas, such optimizations will be incorporated into future iterations of SuperSim, potentially providing major computational complexity and runtime improvements.

\section{Ongoing: Performance Improvements through Parallelization}
\label{sec:ongoing-performance-improvements}
We observe that the following steps in the circuit-cutting process are amenable to parallelization:
\begin{itemize}
    \item Parsing circuits to find cut locations.
    \item Simulating each variant associated with a circuit fragment.
    \item Postprocessing the probability distributions generated by the aforementioned simulations in order to reconstruct the final probability distribution for the original circuit.
\end{itemize}
Parallelizing these operations has been shown to reduce simulation costs for large circuits \cite{perlin2019parallelizing}. We intend to further improve the efficiency of the SuperSim framework by exploring parallelization methods and leveraging the parallel processing power of GPUs and multi-core computing clusters. 


\section{Ongoing: Support of Additional Fragment Evaluators}

SuperSim is currently a classical simulator of quantum circuits based on Cirq and Stim simulation backends. Future versions of SuperSim will support additional fragment evaluation backends, producing composite probability distributions that consist of results from a variety of technologies that include both real QCs and state-of-art classical quantum circuit simulators. SuperSim cuts circuits in order to maximize the size of Clifford circuits and minimize the size of non-Clifford circuits. These fragments are simulated classically, but minor adjustments to our framework would enable the non-Clifford circuits to be run on a real quantum machine. In the future, we envision the development of dynamic techniques that identify circuit fragment and quantum machine / circuit simulator pairings, running fragment variants on the most resource-efficient backend. 

\section{Ongoing: Road Map for Future Evaluation}
\label{sec:ongoing-roadmap-for-future-eval}

The results presented here are a preliminary effort to assess SuperSim's performance in terms of runtime and processing quality in terms of output fidelity. However, they are incomplete, and efforts are in-progress to expand the coverage of SuperSim benchmarking to discover the applications for which it provides significant performance gains for quantum circuit simulation. Our intuition guided us to explore near-Clifford circuits as our first set of experiments, but our results in Section~\ref{sec:results} lead us to believe that we need to additionally search for near-Clifford circuits that are characterized by high entanglement. Further, we need to increase the scale of our evaluation so that we can discover what upper bounds exist in terms of SuperSim's qubit capacity. These next steps are all part of the SuperSim development roadmap.

\section{Conclusion}
\label{sec:conclusion}

Quantum computing 
benefits from quantum circuit simulation for applications such as VQAs and QEC. Exact quantum circuit simulators quickly become intractable to evaluate on classical hardware as a quantum system grows in size. However, there has recently been progress in developing more efficient quantum circuit simulation techniques for near-Clifford quantum circuits. In this industry track paper, we debut the Super.tech SuperSim framework, a new approach for high fidelity and scalable quantum circuit simulation based on Clifford-based simulation and circuit cutting. Our results show that Clifford-based circuit cutting accelerates the simulation of near-Clifford circuits beyond the frontier of alternative state-of-art techniques, allowing 100s of qubits to be evaluated with modest runtimes.

\bibliographystyle{IEEEtranS}
\bibliography{refs}

\begin{thebibliography}{10}
\providecommand{\url}[1]{#1}
\csname url@samestyle\endcsname
\providecommand{\newblock}{\relax}
\providecommand{\bibinfo}[2]{#2}
\providecommand{\BIBentrySTDinterwordspacing}{\spaceskip=0pt\relax}
\providecommand{\BIBentryALTinterwordstretchfactor}{4}
\providecommand{\BIBentryALTinterwordspacing}{\spaceskip=\fontdimen2\font plus
\BIBentryALTinterwordstretchfactor\fontdimen3\font minus
  \fontdimen4\font\relax}
\providecommand{\BIBforeignlanguage}[2]{{%
\expandafter\ifx\csname l@#1\endcsname\relax
\typeout{** WARNING: IEEEtranS.bst: No hyphenation pattern has been}%
\typeout{** loaded for the language `#1'. Using the pattern for}%
\typeout{** the default language instead.}%
\else
\language=\csname l@#1\endcsname
\fi
#2}}
\providecommand{\BIBdecl}{\relax}
\BIBdecl

\bibitem{aaronson2004improved}
S.~Aaronson and D.~Gottesman, ``Improved simulation of stabilizer circuits,''
  \emph{Physical Review A}, vol.~70, no.~5, p. 052328, 2004.

\bibitem{anschuetz2022interpretable}
E.~R. Anschuetz, H.-Y. Hu, J.-L. Huang, and X.~Gao, ``Interpretable quantum
  advantage in neural sequence learning,'' \emph{arXiv preprint
  arXiv:2209.14353}, 2022.

\bibitem{bennink2017unbiased}
R.~S. Bennink, E.~M. Ferragut, T.~S. Humble, J.~A. Laska, J.~J. Nutaro, M.~G.
  Pleszkoch, and R.~C. Pooser, ``Unbiased simulation of near-clifford quantum
  circuits,'' \emph{Physical Review A}, vol.~95, no.~6, p. 062337, 2017.

\bibitem{biamonte2017quantum}
J.~Biamonte, P.~Wittek, N.~Pancotti, P.~Rebentrost, N.~Wiebe, and S.~Lloyd,
  ``Quantum machine learning,'' \emph{Nature}, vol. 549, no. 7671, pp.
  195--202, 2017.

\bibitem{bravyi2019simulation}
S.~Bravyi, D.~Browne, P.~Calpin, E.~Campbell, D.~Gosset, and M.~Howard,
  ``Simulation of quantum circuits by low-rank stabilizer decompositions,''
  \emph{Quantum}, vol.~3, p. 181, 2019.

\bibitem{cerezo2021cost}
M.~Cerezo, A.~Sone, T.~Volkoff, L.~Cincio, and P.~J. Coles, ``Cost function
  dependent barren plateaus in shallow parametrized quantum circuits,''
  \emph{Nature communications}, vol.~12, no.~1, p. 1791, 2021.

\bibitem{cirac2021matrix}
J.~I. Cirac, D.~Perez-Garcia, N.~Schuch, and F.~Verstraete, ``Matrix product
  states and projected entangled pair states: Concepts, symmetries, theorems,''
  \emph{Reviews of Modern Physics}, vol.~93, no.~4, p. 045003, 2021.

\bibitem{czarnik2020error}
P.~Czarnik, A.~Arrasmith, P.~J. Coles, and L.~Cincio, ``Error mitigation with
  clifford quantum-circuit data,'' 2020.

\bibitem{darmawan2017tensor}
A.~S. Darmawan and D.~Poulin, ``Tensor-network simulations of the surface code
  under realistic noise,'' \emph{Physical review letters}, vol. 119, no.~4, p.
  040502, 2017.

\bibitem{das2021adapt}
P.~Das, S.~Tannu, S.~Dangwal, and M.~Qureshi, ``Adapt: Mitigating idling errors
  in qubits via adaptive dynamical decoupling,'' in \emph{MICRO-54: 54th Annual
  IEEE/ACM International Symposium on Microarchitecture}, 2021, pp. 950--962.

\bibitem{dawson2005solovay}
C.~M. Dawson and M.~A. Nielsen, ``The solovay-kitaev algorithm,'' \emph{arXiv
  preprint quant-ph/0505030}, 2005.

\bibitem{de2007massively}
K.~De~Raedt, K.~Michielsen, H.~De~Raedt, B.~Trieu, G.~Arnold, M.~Richter,
  T.~Lippert, H.~Watanabe, and N.~Ito, ``Massively parallel quantum computer
  simulator,'' \emph{Computer Physics Communications}, vol. 176, no.~2, pp.
  121--136, 2007.

\bibitem{cirq_developers_2022_7465577}
\BIBentryALTinterwordspacing
C.~Developers, ``Cirq,'' Dec. 2022, {See full list of authors on Github:
  https://github .com/quantumlib/Cirq/graphs/contributors}. [Online].
  Available: \url{https://doi.org/10.5281/zenodo.7465577}
\BIBentrySTDinterwordspacing

\bibitem{farhi2022quantum}
E.~Farhi, J.~Goldstone, S.~Gutmann, and L.~Zhou, ``The quantum approximate
  optimization algorithm and the sherrington-kirkpatrick model at infinite
  size,'' \emph{Quantum}, vol.~6, p. 759, 2022.

\bibitem{fowler2012surface}
A.~G. Fowler, M.~Mariantoni, J.~M. Martinis, and A.~N. Cleland, ``Surface
  codes: Towards practical large-scale quantum computation,'' \emph{Physical
  Review A}, vol.~86, no.~3, p. 032324, 2012.

\bibitem{gao2022enhancing}
X.~Gao, E.~R. Anschuetz, S.-T. Wang, J.~I. Cirac, and M.~D. Lukin, ``Enhancing
  generative models via quantum correlations,'' \emph{Physical Review X},
  vol.~12, no.~2, p. 021037, 2022.

\bibitem{gidney2021stim}
C.~Gidney, ``Stim: a fast stabilizer circuit simulator,'' \emph{Quantum},
  vol.~5, p. 497, 2021.

\bibitem{gokhale2022supercheq}
P.~Gokhale, E.~Anschuetz, C.~Campbell, F.~Chong, E.~Dahl, P.~Frederick,
  E.~Jones, B.~Hall, S.~Issa, P.~Goiporia \emph{et~al.}, ``Supercheq: Quantum
  advantage for distributed databases,'' \emph{arXiv preprint
  arXiv:2212.03850}, 2022.

\bibitem{gottesman1998heisenberg}
D.~Gottesman, ``The heisenberg representation of quantum computers,''
  \emph{arXiv preprint quant-ph/9807006}, 1998.

\bibitem{gottesman1998theory}
D.~Gottesman, ``Theory of fault-tolerant quantum computation,'' \emph{Physical
  Review A}, vol.~57, no.~1, p. 127, 1998.

\bibitem{gottesman1999demonstrating}
D.~Gottesman and I.~L. Chuang, ``Demonstrating the viability of universal
  quantum computation using teleportation and single-qubit operations,''
  \emph{Nature}, vol. 402, no. 6760, pp. 390--393, 1999.

\bibitem{grurl2022noise}
T.~Grurl, J.~Fu{\ss}, and R.~Wille, ``Noise-aware quantum circuit simulation
  with decision diagrams,'' \emph{IEEE Transactions on Computer-Aided Design of
  Integrated Circuits and Systems}, 2022.

\bibitem{harrow2009quantum}
A.~W. Harrow, A.~Hassidim, and S.~Lloyd, ``Quantum algorithm for linear systems
  of equations,'' \emph{Physical review letters}, vol. 103, no.~15, p. 150502,
  2009.

\bibitem{kandala2017hardware}
A.~Kandala, A.~Mezzacapo, K.~Temme, M.~Takita, M.~Brink, J.~M. Chow, and J.~M.
  Gambetta, ``Hardware-efficient variational quantum eigensolver for small
  molecules and quantum magnets,'' \emph{Nature}, vol. 549, no. 7671, pp.
  242--246, 2017.

\bibitem{kerzner2021clifford}
A.~Kerzner, ``Clifford simulation: Techniques and applications,'' Master's
  thesis, University of Waterloo, 2021.

\bibitem{kissinger2022simulating}
A.~Kissinger and J.~van~de Wetering, ``Simulating quantum circuits with
  zx-calculus reduced stabiliser decompositions,'' \emph{Quantum Science and
  Technology}, 2022.

\bibitem{koh2017computing}
D.~E. Koh, M.~D. Penney, and R.~W. Spekkens, ``Computing quopit clifford
  circuit amplitudes by the sum-over-paths technique,'' \emph{arXiv preprint
  arXiv:1702.03316}, 2017.

\bibitem{liu2022classical}
J.~Liu, A.~Gonzales, and Z.~H. Saleem, ``Classical simulators as quantum error
  mitigators via circuit cutting,'' \emph{arXiv preprint arXiv:2212.07335},
  2022.

\bibitem{liu2021}
\BIBentryALTinterwordspacing
Y.~A. Liu, X.~L. Liu, F.~N. Li, H.~Fu, Y.~Yang, J.~Song, P.~Zhao, Z.~Wang,
  D.~Peng, H.~Chen, C.~Guo, H.~Huang, W.~Wu, and D.~Chen, ``Closing the
  "quantum supremacy" gap: Achieving real-time simulation of a random quantum
  circuit using a new sunway supercomputer,'' in \emph{Proceedings of the
  International Conference for High Performance Computing, Networking, Storage
  and Analysis}, ser. SC '21.\hskip 1em plus 0.5em minus 0.4em\relax New York,
  NY, USA: Association for Computing Machinery, 2021. [Online]. Available:
  \url{https://doi.org/10.1145/3458817.3487399}
\BIBentrySTDinterwordspacing

\bibitem{lykov2022tensor}
D.~Lykov, R.~Schutski, A.~Galda, V.~Vinokur, and Y.~Alexeev, ``Tensor network
  quantum simulator with step-dependent parallelization,'' in \emph{2022 IEEE
  International Conference on Quantum Computing and Engineering (QCE)}.\hskip
  1em plus 0.5em minus 0.4em\relax IEEE, 2022, pp. 582--593.

\bibitem{markov2008simulating}
I.~L. Markov and Y.~Shi, ``Simulating quantum computation by contracting tensor
  networks,'' \emph{SIAM Journal on Computing}, vol.~38, no.~3, pp. 963--981,
  2008.

\bibitem{medvidovic2021classical}
M.~Medvidovi{\'c} and G.~Carleo, ``Classical variational simulation of the
  quantum approximate optimization algorithm,'' \emph{npj Quantum Information},
  vol.~7, no.~1, pp. 1--7, 2021.

\bibitem{moll2018quantum}
N.~Moll, P.~Barkoutsos, L.~S. Bishop, J.~M. Chow, A.~Cross, D.~J. Egger,
  S.~Filipp, A.~Fuhrer, J.~M. Gambetta, M.~Ganzhorn \emph{et~al.}, ``Quantum
  optimization using variational algorithms on near-term quantum devices,''
  \emph{Quantum Science and Technology}, vol.~3, no.~3, p. 030503, 2018.

\bibitem{mike_ike_2020}
M.~A. Nielsen and I.~Chuang, \emph{Quantum computation and quantum
  information}.\hskip 1em plus 0.5em minus 0.4em\relax Cambridge University
  Press, 2010.

\bibitem{o2017quantum}
J.~O'Gorman and E.~T. Campbell, ``Quantum computation with realistic
  magic-state factories,'' \emph{Physical Review A}, vol.~95, no.~3, p. 032338,
  2017.

\bibitem{pashayan2022fast}
H.~Pashayan, O.~Reardon-Smith, K.~Korzekwa, and S.~D. Bartlett, ``Fast
  estimation of outcome probabilities for quantum circuits,'' \emph{PRX
  Quantum}, vol.~3, no.~2, p. 020361, 2022.

\bibitem{pednault2017breaking}
E.~Pednault, J.~A. Gunnels, G.~Nannicini, L.~Horesh, T.~Magerlein,
  E.~Solomonik, and R.~Wisnieff, ``Breaking the 49-qubit barrier in the
  simulation of quantum circuits,'' \emph{arXiv preprint arXiv:1710.05867},
  vol.~15, 2017.

\bibitem{peng2020simulating}
T.~Peng, A.~W. Harrow, M.~Ozols, and X.~Wu, ``Simulating large quantum circuits
  on a small quantum computer,'' \emph{Physical Review Letters}, vol. 125,
  no.~15, p. 150504, 2020.

\bibitem{perlin2019parallelizing}
M.~Perlin, T.~Tomesh, B.~Pearlman, W.~Tang, Y.~Alexeev, and M.~Suchara,
  ``Parallelizing simulations of large quantum circuits,'' 2019.

\bibitem{perlin2021quantum}
M.~A. Perlin, Z.~H. Saleem, M.~Suchara, and J.~C. Osborn, ``Quantum circuit
  cutting with maximum-likelihood tomography,'' \emph{npj Quantum Information},
  vol.~7, no.~1, pp. 1--8, 2021.

\bibitem{peruzzo2014variational}
A.~Peruzzo, J.~McClean, P.~Shadbolt, M.-H. Yung, X.-Q. Zhou, P.~J. Love,
  A.~Aspuru-Guzik, and J.~L. O’brien, ``A variational eigenvalue solver on a
  photonic quantum processor,'' \emph{Nature communications}, vol.~5, no.~1,
  pp. 1--7, 2014.

\bibitem{ravi2022cafqa}
G.~S. Ravi, P.~Gokhale, Y.~Ding, W.~Kirby, K.~Smith, J.~M. Baker, P.~J. Love,
  H.~Hoffmann, K.~R. Brown, and F.~T. Chong, ``Cafqa: A classical simulation
  bootstrap for variational quantum algorithms,'' in \emph{Proceedings of the
  28th ACM International Conference on Architectural Support for Programming
  Languages and Operating Systems, Volume 1}, 2022, pp. 15--29.

\bibitem{QECIntro}
\BIBentryALTinterwordspacing
J.~Roffe, ``Quantum error correction: an introductory guide,''
  \emph{Contemporary Physics}, vol.~60, no.~3, p. 226–245, Jul 2019.
  [Online]. Available: \url{http://dx.doi.org/10.1080/00107514.2019.1667078}
\BIBentrySTDinterwordspacing

\bibitem{schollwock2011density}
U.~Schollw{\"o}ck, ``The density-matrix renormalization group in the age of
  matrix product states,'' \emph{Annals of physics}, vol. 326, no.~1, pp.
  96--192, 2011.

\bibitem{shor1999polynomial}
P.~W. Shor, ``Polynomial-time algorithms for prime factorization and discrete
  logarithms on a quantum computer,'' \emph{SIAM review}, vol.~41, no.~2, pp.
  303--332, 1999.

\bibitem{strikis2021learningbased}
A.~Strikis, D.~Qin, Y.~Chen, S.~C. Benjamin, and Y.~Li, ``Learning-based
  quantum error mitigation,'' 2021.

\bibitem{Qiskit}
A.~tA~v, M.~S. ANIS, Abby-Mitchell, H.~Abraham, AduOffei, R.~Agarwal,
  G.~Agliardi, M.~Aharoni, V.~Ajith, I.~Y. Akhalwaya, G.~Aleksandrowicz,
  T.~Alexander, M.~Amy, S.~Anagolum, Anthony-Gandon, I.~F. Araujo, E.~Arbel,
  A.~Asfaw, A.~Athalye, A.~Avkhadiev, C.~Azaustre, P.~BHOLE, V.~Bajpe,
  A.~Banerjee, S.~Banerjee, W.~Bang, A.~Bansal, P.~Barkoutsos, A.~Barnawal,
  G.~Barron, G.~S. Barron, L.~Bello, Y.~Ben-Haim, M.~C. Bennett, D.~Bevenius,
  D.~Bhatnagar, P.~Bhatnagar, A.~Bhobe, P.~Bianchini, L.~S. Bishop, C.~Blank,
  S.~Bolos, S.~Bopardikar, S.~Bosch, S.~Brandhofer, Brandon, S.~Bravyi,
  Bryce-Fuller, D.~Bucher, L.~Burgholzer, A.~Burov, F.~Cabrera, P.~Calpin,
  L.~Capelluto, J.~Carballo, G.~Carrascal, A.~Carriker, I.~Carvalho,
  R.~Chakrabarti, A.~Chen, C.-F. Chen, E.~Chen, J.~C. Chen, R.~Chen,
  F.~Chevallier, K.~Chinda, R.~Cholarajan, J.~M. Chow, S.~Churchill,
  CisterMoke, C.~Claus, C.~Clauss, C.~Clothier, R.~Cocking, R.~Cocuzzo,
  J.~Connor, F.~Correa, Z.~Crockett, A.~J. Cross, A.~W. Cross, S.~Cross,
  J.~Cruz-Benito, C.~Culver, A.~D. C{\'o}rcoles-Gonzales, N.~D, S.~Dague, T.~E.
  Dandachi, A.~N. Dangwal, J.~Daniel, M.~Daniels, M.~Dartiailh, A.~R. Davila,
  F.~Debouni, A.~Dekusar, A.~Deshmukh, M.~Deshpande, D.~Ding, J.~Doi, E.~M.
  Dow, P.~Downing, E.~Drechsler, M.~S. Drudis, E.~Dumitrescu, K.~Dumon,
  I.~Duran, K.~EL-Safty, E.~Eastman, G.~Eberle, A.~Ebrahimi, P.~Eendebak,
  D.~Egger, ElePT, I.~Elsayed, Emilio, A.~Espiricueta, M.~Everitt, D.~Facoetti,
  Farida, P.~M. Fern{\'a}ndez, S.~Ferracin, D.~Ferrari, A.~H. Ferrera,
  R.~Fouilland, A.~Frisch, A.~Fuhrer, B.~Fuller, M.~GEORGE, J.~Gacon, B.~G.
  Gago, C.~Gambella, J.~M. Gambetta, A.~Gammanpila, L.~Garcia, T.~Garg,
  S.~Garion, J.~R. Garrison, J.~Garrison, T.~Gates, N.~Gavrielov,
  G.~Gentinetta, H.~Georgiev, L.~Gil, A.~Gilliam, A.~Giridharan, Glen,
  J.~Gomez-Mosquera, Gonzalo, S.~de~la Puente~Gonz{\'a}lez, J.~Gorzinski,
  I.~Gould, D.~Greenberg, D.~Grinko, W.~Guan, D.~Guijo,
  Guillermo-Mijares-Vilarino, J.~A. Gunnels, H.~Gupta, N.~Gupta, J.~M.
  G{\"u}nther, M.~Haglund, I.~Haide, I.~Hamamura, O.~C. Hamido, F.~Harkins,
  K.~Hartman, A.~Hasan, V.~Havlicek, J.~Hellmers, {\L}.~Herok, R.~Hill,
  S.~Hillmich, C.~Hong, H.~Horii, C.~Howington, S.~Hu, W.~Hu, C.-H. Huang,
  J.~Huang, R.~Huisman, H.~Imai, T.~Imamichi, K.~Ishizaki, Ishwor, R.~Iten,
  T.~Itoko, A.~Ivrii, A.~Javadi, A.~Javadi-Abhari, W.~Javed, Q.~Jianhua,
  M.~Jivrajani, K.~Johns, S.~Johnstun, Jonathan-Shoemaker, JosDenmark,
  JoshDumo, J.~Judge, T.~Kachmann, A.~Kale, N.~Kanazawa, J.~Kane, Kang-Bae,
  A.~Kapila, A.~Karazeev, P.~Kassebaum, T.~Kehrer, J.~Kelso, S.~Kelso, H.~van
  Kemenade, V.~Khanderao, S.~King, Y.~Kobayashi, Kovi11Day, A.~Kovyrshin,
  R.~Krishnakumar, P.~Krishnamurthy, V.~Krishnan, K.~Krsulich, P.~Kumkar,
  G.~Kus, R.~LaRose, E.~Lacal, R.~Lambert, H.~Landa, J.~Lapeyre, D.~Lasecki,
  J.~Latone, S.~Lawrence, C.~Lee, G.~Li, T.~J. Liang, J.~Lishman, D.~Liu,
  P.~Liu, Lolcroc, A.~K. M, L.~Madden, Y.~Maeng, S.~Maheshkar, K.~Majmudar,
  A.~Malyshev, M.~E. Mandouh, J.~Manela, Manjula, J.~Marecek, M.~Marques,
  K.~Marwaha, D.~Maslov, P.~Maszota, D.~Mathews, A.~Matsuo, F.~Mazhandu,
  D.~McClure, M.~McElaney, J.~McElroy, C.~McGarry, D.~McKay, D.~McPherson,
  S.~Meesala, D.~Meirom, C.~Mendell, T.~Metcalfe, M.~Mevissen, A.~Meyer,
  A.~Mezzacapo, R.~Midha, D.~Millar, D.~Miller, H.~Miller, Z.~Minev,
  A.~Mitchell, N.~Moll, A.~Montanez, G.~Monteiro, M.~D. Mooring, R.~Morales,
  N.~Moran, D.~Morcuende, S.~Mostafa, M.~Motta, R.~Moyard, P.~Murali,
  D.~Murata, J.~M{\"u}ggenburg, T.~NEMOZ, D.~Nadlinger, K.~Nakanishi,
  G.~Nannicini, P.~Nation, E.~Navarro, Y.~Naveh, S.~W. Neagle, P.~Neuweiler,
  A.~Ngoueya, T.~Nguyen, J.~Nicander, Nick-Singstock, P.~Niroula, H.~Norlen,
  NuoWenLei, L.~J. O'Riordan, O.~Ogunbayo, P.~Ollitrault, T.~Onodera,
  R.~Otaolea, S.~Oud, D.~Padilha, H.~Paik, S.~Pal, Y.~Pang, A.~Panigrahi, V.~R.
  Pascuzzi, S.~Perriello, E.~Peterson, A.~Phan, K.~Pilch, F.~Piro, M.~Pistoia,
  C.~Piveteau, J.~Plewa, P.~Pocreau, C.~Possel, A.~Pozas-Kerstjens, R.~Pracht,
  M.~Prokop, V.~Prutyanov, S.~Puri, D.~Puzzuoli, Pythonix, J.~P{\'e}rez,
  Quant02, Quintiii, R.~I. Rahman, A.~Raja, R.~Rajeev, I.~Rajput, N.~Ramagiri,
  A.~Rao, R.~Raymond, O.~Reardon-Smith, R.~M.-C. Redondo, M.~Reuter, J.~Rice,
  M.~Riedemann, Rietesh, D.~Risinger, P.~Rivero, M.~L. Rocca, D.~M.
  Rodr{\'\i}guez, RohithKarur, B.~Rosand, M.~Rossmannek, M.~Ryu, T.~SAPV,
  N.~R.~C. Sa, A.~Saha, A.~Ash-Saki, A.~Salman, S.~Sanand, M.~Sandberg,
  H.~Sandesara, R.~Sapra, H.~Sargsyan, A.~Sarkar, N.~Sathaye, N.~Savola,
  B.~Schmitt, C.~Schnabel, Z.~Schoenfeld, T.~L. Scholten, E.~Schoute,
  M.~Schulterbrandt, J.~Schwarm, P.~Schweigert, J.~Seaward, Sergi, I.~F.
  Sertage, K.~Setia, F.~Shah, N.~Shammah, W.~Shanks, R.~Sharma, P.~Shaw,
  Y.~Shi, J.~Shoemaker, A.~Silva, A.~Simonetto, D.~Singh, D.~Singh, P.~Singh,
  P.~Singkanipa, Y.~Siraichi, Siri, J.~Sistos, J.~Sistos, I.~Sitdikov,
  S.~Sivarajah, Slavikmew, M.~B. Sletfjerding, J.~A. Smolin, M.~Soeken, I.~O.
  Sokolov, I.~Sokolov, V.~P. Soloviev, SooluThomas, Starfish, D.~Steenken,
  M.~Stypulkoski, A.~Suau, S.~Sun, K.~J. Sung, M.~Suwama, O.~S{\l}owik,
  R.~Taeja, H.~Takahashi, T.~Takawale, I.~Tavernelli, C.~Taylor, P.~Taylour,
  S.~Thomas, K.~Tian, M.~Tillet, M.~Tod, M.~Tomasik, C.~Tornow, E.~de~la Torre,
  J.~L.~S. Toural, K.~Trabing, M.~Treinish, D.~Trenev, TrishaPe, F.~Truger,
  TsafrirA, G.~Tsilimigkounakis, D.~Tulsi, D.~Tuna, W.~Turner, Y.~Vaknin, C.~R.
  Valcarce, F.~Varchon, A.~Vartak, A.~C. Vazquez, P.~Vijaywargiya, V.~Villar,
  B.~Vishnu, D.~Vogt-Lee, C.~Vuillot, WQ, J.~Weaver, J.~Weidenfeller,
  R.~Wieczorek, J.~A. Wildstrom, J.~Wilson, E.~Winston, WinterSoldier, J.~J.
  Woehr, S.~Woerner, R.~Woo, C.~J. Wood, R.~Wood, S.~Wood, J.~Wootton,
  M.~Wright, L.~Xing, J.~YU, Yaiza, B.~Yang, U.~Yang, J.~Yao, D.~Yeralin,
  R.~Yonekura, D.~Yonge-Mallo, R.~Yoshida, R.~Young, J.~Yu, L.~Yu,
  Yuma-Nakamura, C.~Zachow, L.~Zdanski, H.~Zhang, E.~Zheltonozhskii, I.~Zidaru,
  B.~Zimmermann, B.~Zindorf, C.~Zoufal, aeddins ibm, alexzhang13, b63, bartek
  bartlomiej, bcamorrison, brandhsn, nick bronn, chetmurthy, choerst ibm,
  comet, dalin27, deeplokhande, dekel.meirom, derwind, dime10, ehchen,
  ewinston, fanizzamarco, fs1132429, gadial, galeinston, georgezhou20, georgios
  ts, gruu, hhorii, hhyap, hykavitha, itoko, jeppevinkel, jessica angel7,
  jezerjojo14, jliu45, johannesgreiner, jscott2, kUmezawa, klinvill,
  krutik2966, ma5x, michelle4654, msuwama, nico lgrs, nrhawkins, ntgiwsvp,
  ordmoj, sagar pahwa, pritamsinha2304, rithikaadiga, ryancocuzzo, saktar unr,
  saswati qiskit, sebastian mair, septembrr, sethmerkel, sg495, shaashwat,
  smturro2, sternparky, strickroman, tigerjack, tsura crisaldo, upsideon,
  vadebayo49, welien, willhbang, wmurphy collabstar, yang.luh, yuri@FreeBSD,
  and M.~{\v{C}}epulkovskis, ``Qiskit: An open-source framework for quantum
  computing,'' 2021.

\bibitem{tang2021cutqc}
W.~Tang, T.~Tomesh, M.~Suchara, J.~Larson, and M.~Martonosi, ``Cutqc: using
  small quantum computers for large quantum circuit evaluations,'' in
  \emph{Proceedings of the 26th ACM International conference on architectural
  support for programming languages and operating systems}, 2021, pp. 473--486.

\bibitem{quantum_ai_team_and_collaborators_2020_4023103}
\BIBentryALTinterwordspacing
Q.~A. team and collaborators, ``qsim,'' Sep. 2020. [Online]. Available:
  \url{https://doi.org/10.5281/zenodo.4023103}
\BIBentrySTDinterwordspacing

\bibitem{superstaq}
S.~D. Team, ``{SuperstaQ}: Connecting applications to quantum hardware,''
  \url{www.super.tech/about-superstaq}, 2021.

\bibitem{tomesh2022supermarq}
T.~Tomesh, P.~Gokhale, V.~Omole, G.~S. Ravi, K.~N. Smith, J.~Viszlai, X.-C. Wu,
  N.~Hardavellas, M.~R. Martonosi, and F.~T. Chong, ``Supermarq: A scalable
  quantum benchmark suite,'' in \emph{2022 IEEE International Symposium on
  High-Performance Computer Architecture (HPCA)}.\hskip 1em plus 0.5em minus
  0.4em\relax IEEE, 2022, pp. 587--603.

\bibitem{uchehara2022rotation}
G.~Uchehara, T.~M. Aamodt, and O.~Di~Matteo, ``Rotation-inspired circuit cut
  optimization,'' \emph{arXiv preprint arXiv:2211.07358}, 2022.

\bibitem{Veitch2014}
\BIBentryALTinterwordspacing
V.~Veitch, S.~A. Hamed~Mousavian, D.~Gottesman, and J.~Emerson, ``The resource
  theory of stabilizer quantum computation,'' \emph{New Journal of Physics},
  vol.~16, no.~1, p. 013009, Jan 2014. [Online]. Available:
  \url{http://dx.doi.org/10.1088/1367-2630/16/1/013009}
\BIBentrySTDinterwordspacing

\bibitem{vidal2003efficient}
G.~Vidal, ``Efficient classical simulation of slightly entangled quantum
  computations,'' \emph{Physical review letters}, vol.~91, no.~14, p. 147902,
  2003.

\bibitem{chongfsqc}
\BIBentryALTinterwordspacing
X.-C. Wu, S.~Di, E.~M. Dasgupta, F.~Cappello, H.~Finkel, Y.~Alexeev, and F.~T.
  Chong, ``Full-state quantum circuit simulation by using data compression,''
  in \emph{Proceedings of the International Conference for High Performance
  Computing, Networking, Storage and Analysis}, ser. SC '19.\hskip 1em plus
  0.5em minus 0.4em\relax New York, NY, USA: Association for Computing
  Machinery, 2019. [Online]. Available:
  \url{https://doi.org/10.1145/3295500.3356155}
\BIBentrySTDinterwordspacing

\bibitem{zhou2020limits}
Y.~Zhou, E.~M. Stoudenmire, and X.~Waintal, ``What limits the simulation of
  quantum computers?'' \emph{Physical Review X}, vol.~10, no.~4, p. 041038,
  2020.

\end{thebibliography}

\end{document}